\documentclass{svjour3}
\usepackage{graphicx}
\usepackage{natbib}

\begin{document}

\title{Noise-Induced Burst and Spike Synchronizations in An Inhibitory Small-World Network of Subthreshold Bursting Neurons}
\author{Sang-Yoon Kim  \and Woochang Lim}

\institute{S.-Y. Kim \at
              Computational Neuroscience Lab., Daegu National University of Education, Daegu 705-115, Korea\\
              \email{sangyoonkim@dnue.ac.kr}
              \and
              W. Lim (corresponding author) \at
              Computational Neuroscience Lab., Daegu National University of Education, Daegu 705-115, Korea\\
              Department of Science Education, Daegu National University of Education, Daegu 705-115, Korea\\
              Tel.: +82-53-620-1348 \\
              Fax: +82-53-620-1525 \\
              \email{woochanglim@dnue.ac.kr}
}

% \date{Received: date / Accepted: date}
% The correct dates will be entered by the editor

\maketitle

\begin{abstract}
For modeling complex synaptic connectivity, we consider the Watts-Strogatz small-world network which interpolates between regular lattice and random network via rewiring, and investigate the effect of small-world connectivity on emergence of noise-induced population synchronization in an inhibitory population of subthreshold bursting Hindmarsh-Rose neurons (which cannot exhibit spontaneous deterministic firings). Thus, noise-induced burst synchronization (synchrony on the slow bursting timescale) and spike synchronization (synchrony on the fast spike timescale) are found to appear in a synchronized region of the $J-D$ plane ($J$: synaptic inhibition strength and $D$: noise intensity). As the rewiring probability $p$ is decreased from 1 (random network) to 0 (regular lattice), the region of spike synchronization shrinks rapidly in the $J-D$ plane, while the region of the burst synchronization decreases slowly. Population synchronization may be well visualized in the raster plot of neural spikes which can be obtained in experiments. Instantaneous population firing rate, $R(t)$, which is directly obtained from the raster plot of spikes, is a realistic population quantity exhibiting collective behaviors with both the slow bursting and the fast spiking timescales. Through frequency filtering, we separate $R(t)$ into $R_b(t)$ (the instantaneous population burst rate describing the slow bursting behavior) and $R_s(t)$ (the instantaneous population spike rate describing the fast intraburst spiking behavior). Then, we characterize the burst and spike synchronization transitions in terms of the bursting and spiking order parameters, based on $R_b(t)$ and $R_s(t)$, respectively. Furthermore, the degree of burst synchronization seen in another raster plot of bursting onset or offset times is well measured in terms of a statistical-mechanical bursting measure $M_b$, introduced by considering the occupation and the pacing patterns of bursting onset or offset times. Similarly, we also measure the degree of the intraburst spike synchronization in terms of a statistical-mechanical spiking measure $M_s$, based on $R_s$. With increase in $p$, both the degrees of the burst and spike synchronizations are found to increase because more long-range connections appear. However, they become saturated for some maximal values of $p$ because long-range short-cuts which appear up to the maximal values of $p$ play sufficient role to get maximal degrees of the burst and spike synchronizations.
\end{abstract}


\maketitle

\section{Introduction}
\label{sec:INT}
Noise-induced firing patterns of subthreshold neurons in the peripheral and central nervous systems have been studied in many physiological and pathophysiological aspects \citep{Braun1}. For example, for encoding environmental electric or thermal stimuli, sensory receptor neurons were found to use the noise-induced firings, which are generated via the ``constructive'' interplay of subthreshold oscillations and noise \citep{Braun2,Longtin1}. In contrast to the suprathreshold case where deterministic firings occur, a distinct characteristic of the noise-induced firings is occurrence of  ``skipping'' of spikes at random integer multiples of a basic oscillation period (i.e., occurrence of stochastic phase locking) \citep{Braun2,Longtin1,Longtin2,Braun1}. This random skipping leads to a multi-modal interspike interval histogram. These noise-induced firings of a single subthreshold neuron become most coherent at an optimal noise intensity, which is called coherence resonance (or autonomous stochastic resonance without periodic forcing) \citep{CR1}. Furthermore, array-enhanced coherence resonance was found to occur via noise-induced synchronization in a population of subthreshold spiking neurons \citep{CR2,CR3,CR4,CR5,CR6}. Here, we are interested in synchronization of noise-induced firings in an ensemble of subthreshold bursting neurons. Bursting occurs when neuronal activity alternates, on a slow timescale, between a silent phase and an active (bursting) phase of fast repetitive spikings \citep{Rinzel1,Rinzel2,Burst1,Burst2,Burst3}. Hence, bursting neurons exhibit two different patterns of synchronization due to the slow and fast timescales of bursting activity. Burst synchronization (synchrony on the slow bursting timescale) refers to a temporal coherence between the active phase onset or offset times of bursting neurons, while spike synchronization (synchrony on the fast spike timescale) characterizes a temporal coherence between intraburst spikes fired by bursting neurons in their respective active phases \citep{Burstsync1,Burstsync2}. Recently, the burst and spike synchronizations have been studied in many aspects \citep{BSsync2,BSsync3,BSsync4,BSsync9,BSsync7,BSsync5,BSsync10,BSsync8,BSsync1,BSsync6,BSsync11,BSsync12}. However, most of these studies were focused on the suprathreshold case where bursting neurons fire deterministic firings, in contrast to subthreshold case of our concern.

In this paper, we study the effect of network architecture on noise-induced burst and spike synchronizations of subthreshold bursting Hindmarsh-Rose (HR) neurons. The conventional Erd\"{o}s-Renyi random graph has been often used for modeling complex connectivity occurring in diverse fields such as social, biological, and technological networks \citep{ER}. Hence, we first consider a random graph of subthreshold bursting HR neurons, and investigate occurrence of the noise-induced population synchronization by varying the synaptic inhibition strength $J$ and the noise intensity $D$. Thus, noise-induced burst and spike synchronizations are found to appear in a synchronous region of the $J-D$ plane. For the random networks, global efficiency of information transfer becomes high because the average path length (i.e., typical separation between two neurons along the minimal path) is short due to long-range connections \citep{Eff1,Eff2}. On the other hand, random networks have poor clustering (i.e., low cliquishness of a typical neighborhood) \citep{Buz2,Sporns}. However, in a real neural network, synaptic connections are known to have complex topology which is neither regular nor random \citep{CN6,Buz2,CN1,CN2,CN3,CN7,CN4,CN5,Sporns}. Hence, we consider the Watts-Strogatz small-world network of subthreshold bursting HR neurons which interpolates between regular lattice (with high clustering) and random network (with short path length) via rewiring \citep{SWN1,SWN2,SWN3}. The Watts-Strogatz model can be regarded as a cluster-friendly extension of the random network by reconciling the six degrees of separation (small-worldness) \citep{SDS1,SDS2} with the circle of friends (clustering). These small-world networks (with predominantly local connections and rare long-distance connections) have been employed in many recent works on various subjects of neurodynamics \citep{CN6,SW2,SW3,SW4,SW5,SW6,SW7,SW8,SW9,SW10,SW11,SW12,SW13}. By varying the rewiring probability $p$ [$p=1$ (0) corresponds to a random network (regular lattice)], we investigate the effect of small-world connectivity on emergence of noise-induced burst and spike synchronizations. As $p$ is decreased from 1, the region of fast spike synchronization shrinks rapidly in the $J-D$ plane, while the region of the slow burst synchronization decreases slowly. Hence, complete synchronization (including both the burst and spike synchronizations) may occur only for sufficiently large $p$ where global effective communication (between distant neurons) for fast spike synchronization may be available via short synaptic paths. On the other hand, for small $p$ only the slow burst synchronization (without spike synchronization) occurs.

These noise-induced burst and spike synchronizations may be well visualized in the raster plot of neural spikes which can be obtained in experiments. Instantaneous population firing rate (IPFR), $R(t)$, which is directly obtained from the raster plot of spikes, is a realistic collective quantity describing population behaviors with both the slow bursting and the fast spiking timescales. Through frequency filtering, we separate $R(t)$ into $R_b(t)$ (the instantaneous population burst rate (IPBR) describing the slow bursting behavior) and $R_s(t)$ (the instantaneous population spike rate (IPSR) describing the fast intraburst spiking behavior). The time-averaged fluctuations of $R_b$ and $R_s$ play the role of bursting and spiking order parameters, ${\cal {O}}_b$ and ${\cal {O}}_s$, used for characterizing the burst and spike synchronization transitions, respectively \citep{Kim1}. By varying $D$, we investigate the noise-induced bursting transition in terms of ${\cal {O}}_b$ for a given $J$, and find that, with increasing the rewiring probability $p$ from 0 (regular lattice) long-range connections begin to appear, and hence the burst-synchronized range of $D$ increases gradually. For fixed $J$ and $D$, we also study the noise-induced spiking transition in terms of ${\cal {O}}_s$ by changing $p$. When passing a critical value $p^*_c$, a transition to spike synchronization is found to occur in small-world networks, because sufficient number of long-range connections 
for occurrence of fast spike synchronization appear. We also consider another raster plot of active phase (bursting) onset or offset times for more direct visualization of bursting behavior. From this type of raster plot, we can directly obtain the IPBR, $R_b^{(on)}(t)$ or $R_b^{(off)}(t)$, without frequency filtering. Then, the time-averaged fluctuations of $R_b^{(on)}(t)$ and $R_b^{(off)}(t)$ also play the role of another bursting order parameters, ${\cal {O}}_b^{(on)}$ and ${\cal {O}}_b^{(off)}$, for the bursting transition \citep{Kim1}. Moreover, the degree of noise-induced burst synchronization seen in the raster plot of bursting onset or offset times is measured in terms of a statistical-mechanical bursting measure $M_b$, which was introduced by considering the occupation and the pacing patterns of bursting onset or offset times in the raster plot \citep{Kim1}. In a similar way, we also employ a statistical-mechanical spiking measure $M_s$, based on $R_s$, and quantitatively measure the degree of the noise-induced intraburst spike synchronization \citep{Kim1}. As $p$ is increased, both the degrees of the noise-induced burst and spike synchronizations become higher because more long-range connections appear. However, the degrees of the burst and spike synchronizations become saturated for some maximal values of $p$ because long-range short-cuts which appear up to the maximal values of $p$ play sufficient role to get their maximal degrees.

This paper is organized as follows. In Sec.~\ref{sec:HR}, we describe an inhibitory population of subthreshold HR neurons. The HR neurons are representative bursting neurons, and they interact through inhibitory GABAergic synapses (involving the $\rm {GABA_A}$ receptors). In Sec.~\ref{sec:SW}, we separate the slow bursting and the fast spiking timescales of the bursting activity, and investigate the effect of the small-world connectivity on the noise-induced burst and spike synchronizations by varying the rewiring probability $p$. These burst and spike synchronization transitions are characterized in terms of the bursting and spiking order parameters, and their degrees are measured by employing statistical-mechanical bursting and spiking measures, respectively. Finally, a summary is given in Section \ref{sec:SUM}.

\section{Inhibitory Population of Subthreshold Bursting Hindmarsh-Rose Neurons}
\label{sec:HR}
We consider an inhibitory population of $N$ subthreshold bursting neurons. As an element in our coupled neural system, we choose the representative bursting HR neuron model which was originally introduced to describe the time evolution of the membrane potential for the pond snails
\citep{HR1,HR2,HR3}. The population dynamics in this neural network is governed by the following set of ordinary differential equations:
\begin{eqnarray}
\frac{dx_i}{dt} &=& y_i - a x^{3}_{i} + b x^{2}_{i} - z_i +I_{DC} +D \xi_{i} -I_{syn,i}, \label{eq:CHRA} \\
\frac{dy_i}{dt} &=& c - d x^{2}_{i} - y_i, \label{eq:CHRB} \\
\frac{dz_i}{dt} &=& r \left[ s (x_i - x_o) - z_i \right], \label{eq:CHRC} \\
\frac{dg_i}{dt}&=& \alpha g_{\infty}(x_i) (1-g_i) - \beta g_i, \;\;\; i=1, \cdots, N, \label{eq:CHRD}
\end{eqnarray}
where
\begin{eqnarray}
I_{syn,i} &=& \frac{J}{d_i^{in}} \sum_{j(\ne i)}^N w_{ij}g_j(t) (x_i - X_{syn}), \label{eq:CHRE} \\
g_{\infty} (x_i) &=& 1/[1+e^{-(x_i-x^*_s)\delta}]. \label{eq:CHRF}
\end{eqnarray}
Here, the state of the $i$th neuron at a time $t$ (measured in units of milliseconds) is characterized by four state variables: the fast membrane potential $x_i$, the fast recovery current $y_i,$ the slow adaptation current $z_i$, and the synaptic gate variable $g_i$ denoting the fraction of open synaptic ion channels. The parameters in the single HR neuron are taken as $a=1.0,$ $b=3.0,$ $c=1.0,$ $d=5.0,$ $r=0.001,$ $s=4.0,$  and $x_o=-1.6$ \citep{Longtin2}.

\begin{figure}
\includegraphics[width=0.7\columnwidth]{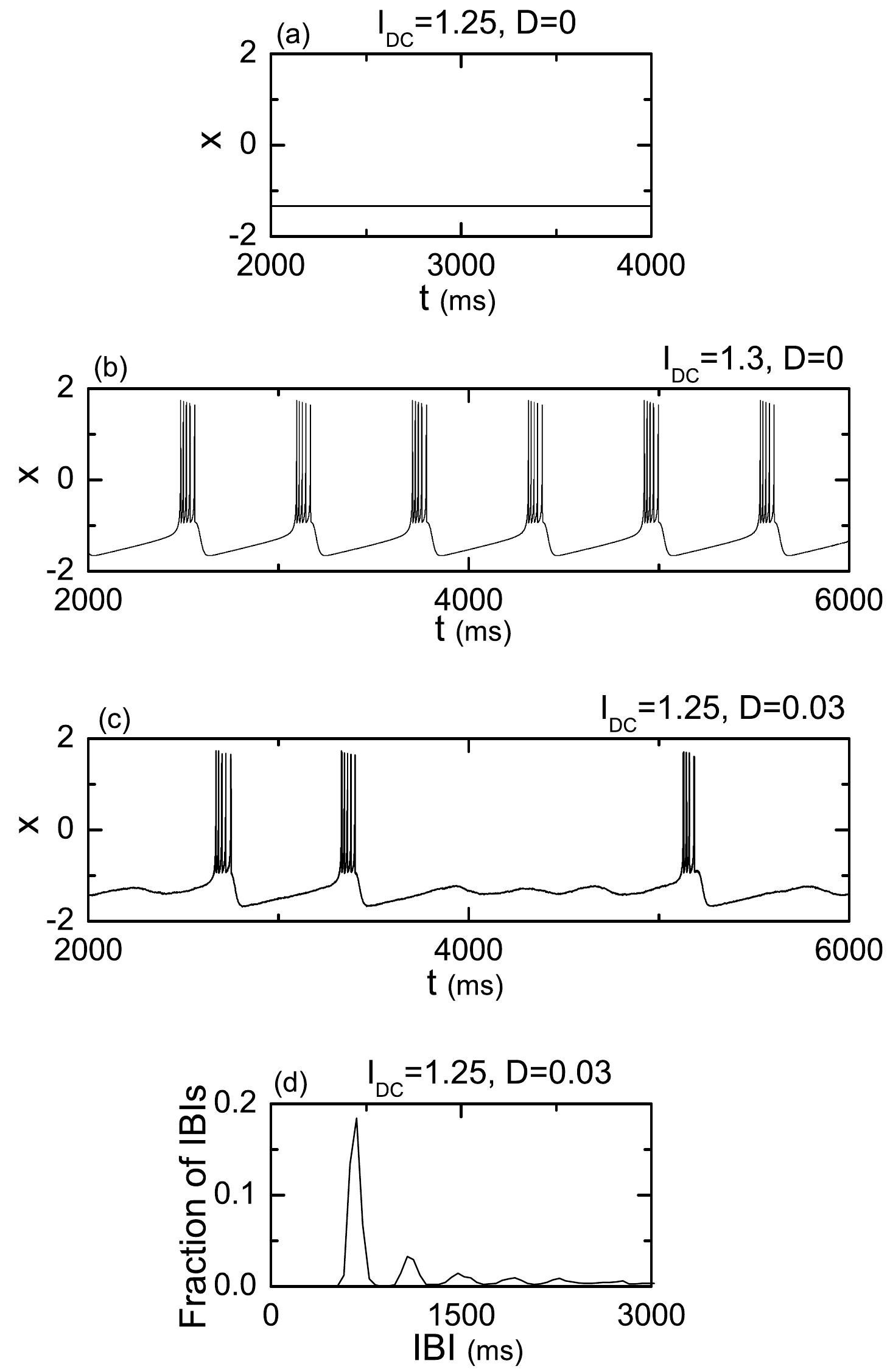}
\caption{Single bursting HR neuron. Time series of the fast membrane potential $x$ for (a) the subthreshold case of $I_{DC}=1.25$ and (b) the suprathreshold case of $I_{DC}=1.3$ in the absence of noise. The dotted horizontal line ($x^*_b=-1$) represents the bursting threshold (the solid and open circles denote the active phase onset and offset times, respectively), while the dashed horizontal line ($x^*_s=0$) represents the spiking threshold within the active phase. (c) Noise-induced intermittent bursting and (d) multi-peaked interburst interval (IBI) histogram for $D=0.03$ in the subthreshold case of $I_{DC}=1.25$. The IBI histogram is made of $5 \times 10^4$ IBIs and the bin size is 50 ms.
}
\label{fig:Single}
\end{figure}

Each bursting HR neuron is stimulated by using the common DC current $I_{DC}$ and an independent Gaussian white noise $\xi_i$ [see the 5th and the 6th terms in Eq.~(\ref{eq:CHRA})] satisfying $\langle \xi_i(t) \rangle =0$ and $\langle \xi_i(t)~\xi_j(t') \rangle = \delta_{ij}~\delta(t-t')$, where
$\langle\cdots\rangle$ denotes the ensemble average. The noise $\xi$ is a parametric one that randomly perturbs the strength of the applied current
$I_{DC}$, and its intensity is controlled by using the parameter $D$. As $I_{DC}$ passes a threshold $I_{DC}^* (\simeq 1.26)$ in the absence of noise (i.e., $D=0$), each single HR neuron exhibits a transition from a resting state [Fig.~\ref{fig:Single}(a)] to a bursting state [Fig.~\ref{fig:Single}(b)]. For the suprathreshold case of $I_{DC}=1.3$, deterministic bursting occurs when neuronal activity alternates, on a slow timescale $(\simeq 609$ ms), between a silent phase and an active (bursting) phase of fast repetitive spikings. An active phase of the bursting activity begins (ends) at a bursting onset (offset) time when the membrane potential $x$ of the bursting HR neuron passes the bursting threshold of $x^*_b=-1$ from below (above). In Fig.~\ref{fig:Single}(b), the dotted horizontal line ($x^*_b=-1$) denotes the bursting threshold (the solid and open circles denote the active phase onset and offset times, respectively), while the dashed horizontal line ($x^*_s=0$) represents the spiking threshold within the active phase. Throughout this paper, we consider the subthreshold case of $I_{DC}=1.25$ where each HR neuron cannot exhibit spontaneous bursting activity without noise. For $D=0.03$, the subthreshold HR neurons show intermittent noise-induced burstings, as shown in Fig.~\ref{fig:Single}(c). This random skipping of bursts occurs roughly at random multiples of a slow timescale of bursting for the noisy HR neuron. However, the slow timescale for the subthreshold spike-driven bursting HR neuron is not defined clearly because the HR neuron model does not have a deterministic slow subsystem which can oscillate in the absence of spikes \citep{Longtin2}. To confirm this random burst skipping, we collect $5 \times 10^4$ interburst intervals (IBIs) from the single HR neuron, where IBIs of an $i$th bursting neuron are referred to intervals between the bursting onset times at which the membrane potential $x_i$ passes a bursting threshold of $x^*_b=-1$ from below. Thus, we get the multi-modal IBI histogram, as shown in Fig.~\ref{fig:Single}(d): the 1st peak occurs at $t=675$ ms and the higher $n$th-order ($n$=2,3,4, ...) peaks seem to appear at $t \simeq 675 + 400~(n-1)$ ms.

The last term in Eq.~(\ref{eq:CHRA}) represents the synaptic coupling of the network. $I_{syn,i}$ of Eq.~(\ref{eq:CHRE}) represents a synaptic current injected into the $i$th neuron. The synaptic connectivity is given by the connection weight matrix $W$ (=$\{ w_{ij} \}$) where  $w_{ij}=1$ if the neuron $j$ is presynaptic to the neuron $i$; otherwise, $w_{ij}=0$. Here, the synaptic connection is modeled by using both the conventional
Erd\"{o}s-Renyi random graph and the Watts-Strogatz small-world network. Then, the in-degree of the $i$th neuron, $d_i^{in}$ (i.e., the number of synaptic inputs to the neuron $i$) is given by $d_i^{in} = \sum_{j(\ne i)}^N w_{ij}$. Here the coupling strength is controlled by the parameter $J$ and $X_{syn}$ is the synaptic reversal potential. Here, we use $X_{syn}=-2$ for the inhibitory synapse. The synaptic gate variable $g$ obeys the 1st order kinetics of Eq.~(\ref{eq:CHRD}) \citep{KE1,KE2}. Here, the normalized concentration of synaptic transmitters, activating the synapse, is assumed to be an instantaneous sigmoidal function of the membrane potential with a spiking threshold $x^*_s$ in Eq.~(\ref{eq:CHRF}), where we set $x^*_s=0$ and $\delta=30$ \citep{HR4}. The transmitter release occurs only when the neuron emits a spike (i.e., its potential $x$ is larger than $x^*_s$). For the inhibitory GABAergic synapse (involving the $\rm{GABA_A}$ receptors), the synaptic channel opening rate, corresponding to the inverse of the synaptic rise time $\tau_r$, is $\alpha=10$ ${\rm ms}^{-1}$, and the synaptic closing rate $\beta$, which is the inverse of the synaptic decay time $\tau_d$, is $\beta=0.1$ ${\rm ms}^{-1}$ \citep{GABA1,GABA2}. Hence, $I_{syn}$ rises fast and decays slowly.

Numerical integration of Eqs.~(\ref{eq:CHRA})-(\ref{eq:CHRD}) is done using the Heun method \citep{SDE} (with the time step $\Delta t=0.01$ ms).
For each realization of the stochastic process, we choose a random initial point $[x_i(0),y_i(0),z_i(0),g_i(0)]$ for the $i$th $(i=1,\dots, N)$ neuron with uniform probability in the range of $x_i(0) \in (-1.7,-1.3)$, $y_i(0) \in (-13,-8)$, $z_i(0) \in (1.0,1.4)$, and $g_i(0) \in (0,0.1)$.

\section{Effect of Small-World Connectivity on Noise-Induced Burst and Spike Synchronizations}
\label{sec:SW}
In this section, we study the effect of small-world connectivity on noise-induced population synchronization in an inhibitory Watts-Strogatz small-world network of subthreshold bursting HR neurons which interpolates between regular lattice and random network via rewiring.
Emergence of noise-induced burst and spike synchronizations is investigated in the $J-D$ plane ($J$: synaptic inhibition strength and $D$: noise intensity) for different values of the rewiring probability $p$. It is thus found that complete noise-induced synchronization (including both the burst and spike synchronizations) occurs for large $p$, while for small $p$ only the noise-induced burst synchronization emerges because more long-range connections are necessary for fast spike synchronization. Through separation of the slow bursting timescale and the fast spiking timescale via frequency filtering, we decompose the IPFR $R(t)$ into the IPBR $R_b(t)$ and the IPSR $R_s(t)$, and characterize the noise-induced bursting and spiking transitions in terms of the bursting and spiking order parameters, ${\cal {O}}_b$ and ${\cal {O}}_s$, based on $R_b(t)$ and $R_s(t)$, respectively. Furthermore, we also measure the degrees of both the noise-induced burst and spike synchronizations by employing statistical-mechanical bursting and spiking measures $M_b$ and $M_s$, respectively and find that their degrees increase with increasing $p$ because more long-range connections appear. However, the degrees of the burst and spike synchronizations become saturated for some maximal values of $p$ because long-range short-cuts which appear up to the maximal values of $p$ play sufficient role to get their maximal degrees.

\begin{figure}
\includegraphics[width=0.7\columnwidth]{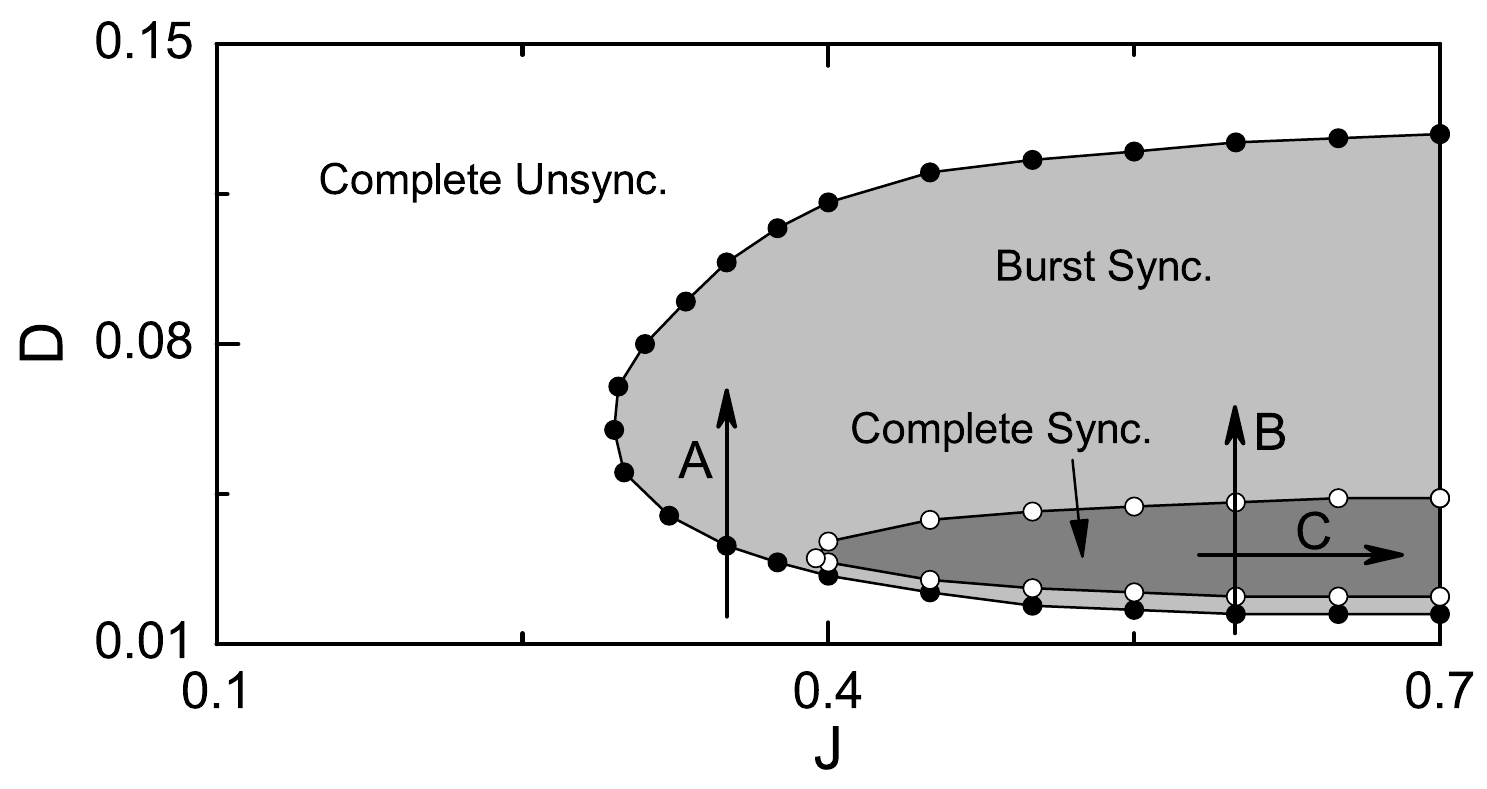}
\caption{State diagram in the $J-D$ plane in the sparse Erd\"{o}s-Renyi random graph of $N$ $(=10^3)$ inhibitory subthreshold bursting HR neurons for $I_{DC}=1.25$ and $M_{syn}=100$. Complete synchronization (including both the burst and spike synchronizations) occurs in the dark gray region, while
in the gray region only the burst synchronization appears. Change in population states along the routes ``A'' and ``B'' and change in bursting type along the route ``C'' are given in Fig.~\ref{fig:PIB}.
}
\label{fig:SD1}
\end{figure}

We first consider the conventional Erd\"{o}s-Renyi random graph of $N$ sparsely-connected bursting HR neurons equidistantly placed on a one-dimensional ring of radius $N/ 2 \pi$. The HR neurons are subthreshold ones which can fire only with the aid of noise, and they
are coupled via inhibitory synapses. A postsynaptic neuron $i$ receives a synaptic input from another presynaptic neuron $j$ with a connection probability $P_{syn}$ $(=M_{syn}/N)$, where $M_{syn}$ is the average number of synaptic inputs per neuron (i.e., $M_{syn} = \langle d_i \rangle$; $d_i$ is the number of synaptic inputs to the neuron $i$ and $\langle \cdots \rangle$ denotes an ensemble-average over all neurons). Here, we consider a sparse case of $M_{syn}=100$. By varying the synaptic inhibition strength $J$ and the noise intensity $D$, we investigate occurrence of noise-induced population synchronization. Figure \ref{fig:SD1} shows the state diagram in the $J-D$ plane. Complete synchronization (including both the burst and spike synchronizations) occurs in the dark gray region, while in the gray region only the burst synchronization (without spike synchronization) appears. For $J < J^*_1$ $(\simeq 0.295)$, no population synchronization occurs. For $J^*_1 < J < J^*_2$ $(\simeq 0.394)$, only slow burst synchronization appears in the gray region, while fast spike synchronization emerges in the dark-gray region for $J> J^*_2$ in addition to the burst synchronization.

\begin{figure}
\includegraphics[width=0.8\columnwidth]{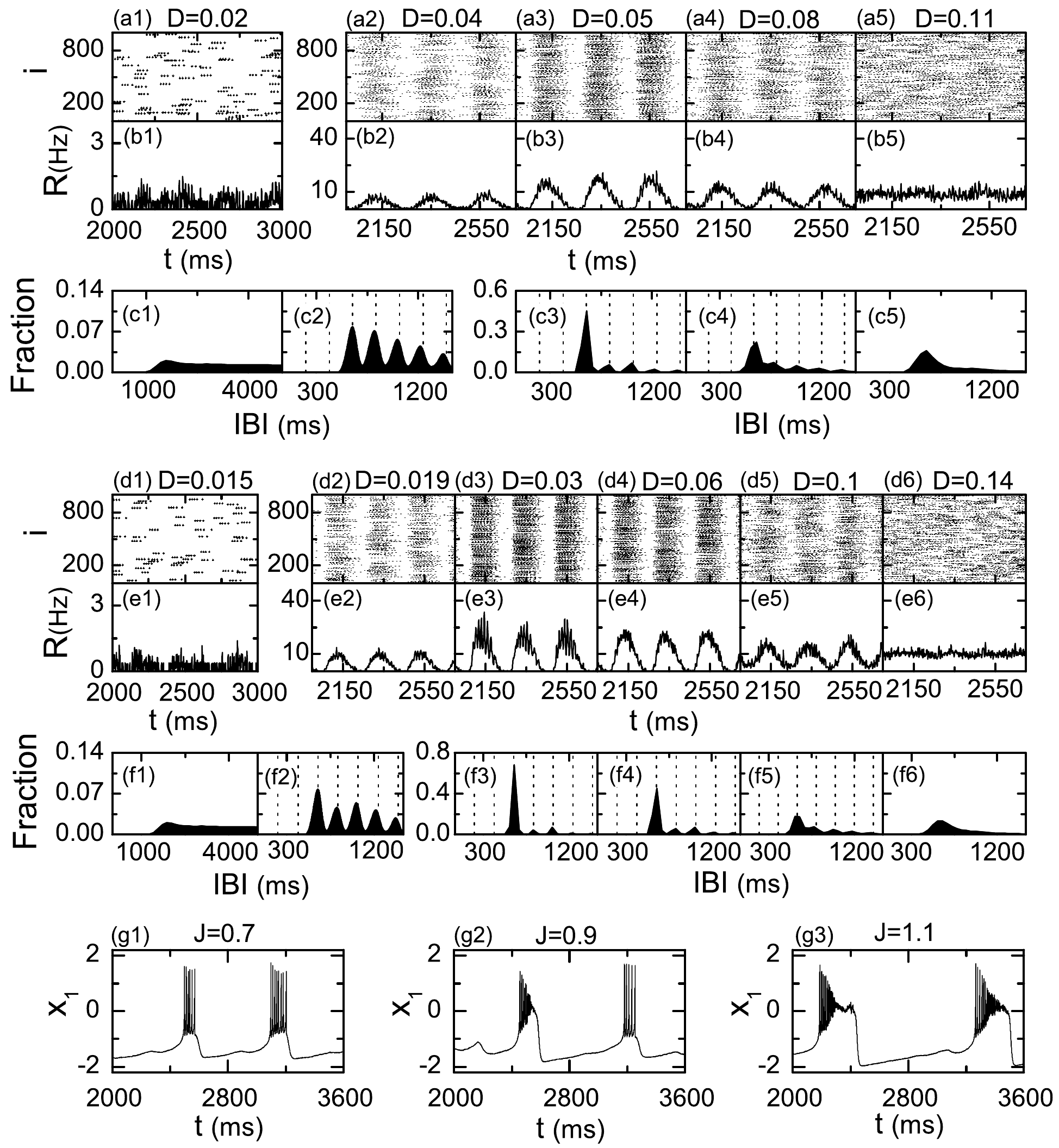}
\caption{Population and individual behaviors along the routes ``A'' and ``B'' in Fig.~\ref{fig:SD1} in the Erd\"{o}s-Renyi random graph of $N$ $(=10^3)$ inhibitory subthreshold bursting HR neurons for $I_{DC}=1.25$ and $M_{syn}=100$. Route ``A'' for $J=0.35$: unsynchronization for $D=0.02$ $\rightarrow$ burst synchronization for $D=0.04$ $\rightarrow$ burst synchronization for $D=0.05$ $\rightarrow$ burst synchronization for $D=0.08$ $\rightarrow$ unsynchronization for $D=0.11$; (a1)-(a5) raster plots of spikes, (b1)-(b5) IPFR kernel estimates $R(t)$, and (c1)-(c5) IBI histograms. Route ``B'' for $J=0.6$: unsynchronization for $D=0.015$ $\rightarrow$ burst synchronization for $D=0.019$ $\rightarrow$ complete synchronization (including both the burst and spike synchronizations) for $D=0.03$ $\rightarrow$ burst synchronization for $D=0.06$ $\rightarrow$ burst synchronization for $D=0.1$ $\rightarrow$ unsynchronization for $D=0.14$; (d1)-(d6) raster plots of spikes, (e1)-(e6) IPFR kernel estimates $R(t)$, and (f1)-(f6) IBI histograms. Change in the bursting type along the route ``C'' for $D=0.03$ in Fig.~\ref{fig:SD1}: (g1) fold-homoclinic (square-wave) bursting for $J=0.7$ $\rightarrow$ (g2) mixed type of fold-homoclinic and fold-Hopf (tapering) burstings for $J=0.9$ $\rightarrow$ (g3) fold-Hopf (tapering) burstings for $J=1.1$. The band width $h$ of the Gaussian kernel function is 1 ms for the IPFR kernel estimate $R(t)$. The IBI histogram is made of $5 \times 10^4$ IBIs, the bin size is 50 ms, and the vertical dotted lines represent the integer multiples of the slow bursting
timescale (i.e., bursting period) $\tau_b$ of $R(t)$: (c2) 208 ms, (c3) 207 ms, (c4) 201 ms, (f2) 208 ms, (f3) 207 ms, (f4) 207 ms, and (f5) 203 ms.
}
\label{fig:PIB}
\end{figure}

Population and individual behaviors along the route ``A'' for $J=0.35$ in Fig.~\ref{fig:SD1} are given in Fig.~\ref{fig:PIB}. The noise-induced burst and spike synchronizations may be well visualized in the raster plot of neural spikes which is a collection of spike trains of individual neurons. Such raster plots of spikes are fundamental data in experimental neuroscience. For describing emergence of population synchronization, we use an experimentally-obtainable IPFR which is often used as a collective quantity showing population behaviors \citep{Wang,Brunel}. The IPFR is directly obtained from the raster plot of neural spikes. To obtain a smooth IPFR from the raster plot of spikes, we employ the kernel density estimation (kernel smoother) \citep{Kernel}. Each spike in the raster plot is convoluted (or blurred) with a kernel function $K_h(t)$ to obtain a smooth estimate of IPFR, $R(t)$:
\begin{equation}
R(t) = \frac{1}{N} \sum_{i=1}^{N} \sum_{s=1}^{n_i} K_h (t-t_{s}^{(i)}),
\label{eq:IPSRK}
\end{equation}
where $t_{s}^{(i)}$ is the $s$th spiking time of the $i$th neuron, $n_i$ is the total number of spikes for the $i$th neuron, and we use a Gaussian
kernel function of band width $h$:
\begin{equation}
K_h (t) = \frac{1}{\sqrt{2\pi}h} e^{-t^2 / 2h^2}, ~~~~ -\infty < t < \infty.
\label{eq:Gaussian}
\end{equation}
For the synchronous case, ``bands" (composed of spikes and indicating population synchronization) are found to be formed in the raster plot. Hence, for a synchronous case, an oscillating IPFR appears, while for an unsynchronized case the IPFR is nearly stationary.
Throughout this study, we consider the population behaviors after the transient time of $2 \times 10^3$ ms. As examples of population states, Figs.~\ref{fig:PIB}(a1)-\ref{fig:PIB}(a5) and Figs.~\ref{fig:PIB}(b1)-\ref{fig:PIB}(b5) show the raster plots of  spikes and the corresponding IPFR kernel estimates $R(t)$ for various values of noise intensity $D$ along the route ``A'' for $J=0.35$. For small $D$, unsynchronized states exist, as shown in the case of $D=0.02$. For this case of unsynchronization sparse spikes are completely scattered in the raster plot of Fig.~\ref{fig:PIB}(a1) and hence the IPFR kernel estimate $R(t)$ in Fig.~\ref{fig:PIB}(b1) is nearly stationary. However, as $D$ passes a lower threshold $D$ $(\simeq 0.033)$, a transition to burst synchronization occurs due to the constructive role of noise to stimulate population synchronization between noise-induced spikes. As an example, see the case of $D=0.04$ where ``bursting bands'' appear successively at nearly regular time intervals [i.e., the slow bursting timescale $\tau_b$ $(\simeq 208$ ms)] in the raster plot of spikes, as shown in Fig.~\ref{fig:PIB}(a2). Within each burst band, spikes are completely scattered, and hence no fast spike synchronization occurs. Consequently, only the slow burst synchronization (without intraburst spike synchronization) emerges. For this case of burst synchronization, the IPFR kernel estimate $R(t)$ in Fig.~\ref{fig:PIB}(b2) shows a slow-wave oscillation with the bursting frequency $f_b$ $\simeq 4.8$ Hz. As $D$ is increased, the smearing degree of the bursting bands becomes reduced, while the density of the bursting bands increases because of the increased bursting rate of the HR neurons, as shown in Fig.~\ref{fig:PIB}(a3) for $D=0.05$. As a result, the amplitude of the slow wave exhibited by the IPFR kernel estimate $R(t)$ increases [see Fig.~\ref{fig:PIB}(b3)]. However, with further increase in $D$, the smearing degree of the bursting bands begins to increase, while the density of the bursting bands decreases because of the reduced bursting rate of the the HR neurons [e.g., see the case of $D=0.08$ in Fig.~\ref{fig:PIB}(a4)]. Consequently, the amplitude of the slow wave shown by the IPFR kernel estimate $R(t)$ decreases, as shown in Fig.~\ref{fig:PIB}(b4). Eventually, when passing a higher threshold $D$ $(\simeq 0.099)$ the smeared bursting bands begin to overlap, and a transition to unsynchronization occurs because of the destructive role of noise to spoil population synchronization between noise-induced spikes. As an example of the unsynchronized state, see the case of $D=0.11$ where the spikes in the raster plot of Fig.~\ref{fig:PIB}(a5) are completely scattered without forming any bursting bands and the IPFR kernel estimate $R(t)$ in Fig.~\ref{fig:PIB}(b5) becomes nearly stationary. Depending on whether the population states are synchronous or unsynchronous, the bursting patterns of individual HR neurons become distinctly different.  To obtain the IBI histograms, we collect $5 \times 10^4$ IBIs from all individual HR neurons. Figures \ref{fig:PIB}(c1)-\ref{fig:PIB}(c5) show the IBI histograms for various values of $D$. For the unsynchronized case of $D=0.02$, the IBI histogram in Fig.~\ref{fig:PIB}(c1) shows a broad distribution with a long tail, and hence the average value of the IBIs ($\simeq 23,947$ ms) becomes very large. However, when passing the lower threshold $D$ $(\simeq 0.033)$, a burst synchronization occurs, and hence a slow-wave oscillation appears in the IPFR kernel estimate $R(t)$. Then, individual HR neurons exhibit intermittent burstings phase-locked to $R(t)$ at random multiples of the slow-wave bursting period $\tau_b$ $(\simeq 208$ ms) of $R(t)$. This random burst skipping (arising from the random phase locking) leads to a multi-modal IBI histogram, as shown in Fig.~\ref{fig:PIB}(c2) for $D=0.04$. The 1st peak in the IBI histogram appears at 3 $\tau_b$ (not $\tau_b$). Hence, individual HR neurons fire sparse burstings mostly every 3rd bursting cycle of $R(t)$. As $D$ is increased, the degree of burst synchronization increases [e.g., see in Figs.~\ref{fig:PIB}(a3) and \ref{fig:PIB}(b3) for $D=0.05$]. For this case, the 1st peak becomes prominently dominant, as shown in Fig.~\ref{fig:PIB}(c3), and hence the tendency of exhibiting burstings every 3rd bursting cycle becomes intensified. However, with further increase in $D$, the heights of peaks are decreased, but their widths are widened. Thus, peaks begin to merge, as shown in Fig.~\ref{fig:PIB}(c4) for $D=0.08$. This merging of peaks results in smearing of bursting bands, and hence the degree of burst synchronization begin to decrease [see Figs.~\ref{fig:PIB}(a4) and \ref{fig:PIB}(b4)]. Eventually, as $D$ passes a higher threshold $(\simeq 0.099)$, unsynchronized states appear (i.e., $R(t)$ becomes nearly stationary), and then the multi-modal structure in the IBI histogram disappears [e.g., see Fig.~\ref{fig:PIB}(c5) for $D=0.11$]. In this way, the IBI histograms have multi-peaked structures due to random burst skipping for the case of burst synchronization, while such peaks disappear in the case of unsynchronization. Similar skipping of spikings (characterized with multi-peaked interspike interval histograms) were also found in inhibitory population of subthreshold spiking neurons \citep{Kim2}. This kind of random burst/spike skipping in networks of inhibitory subthreshold bursting/spiking neurons is a collective effect because it occurs due to a driving by a coherent ensemble-averaged synaptic current.

\begin{figure}
\includegraphics[width=0.7\columnwidth]{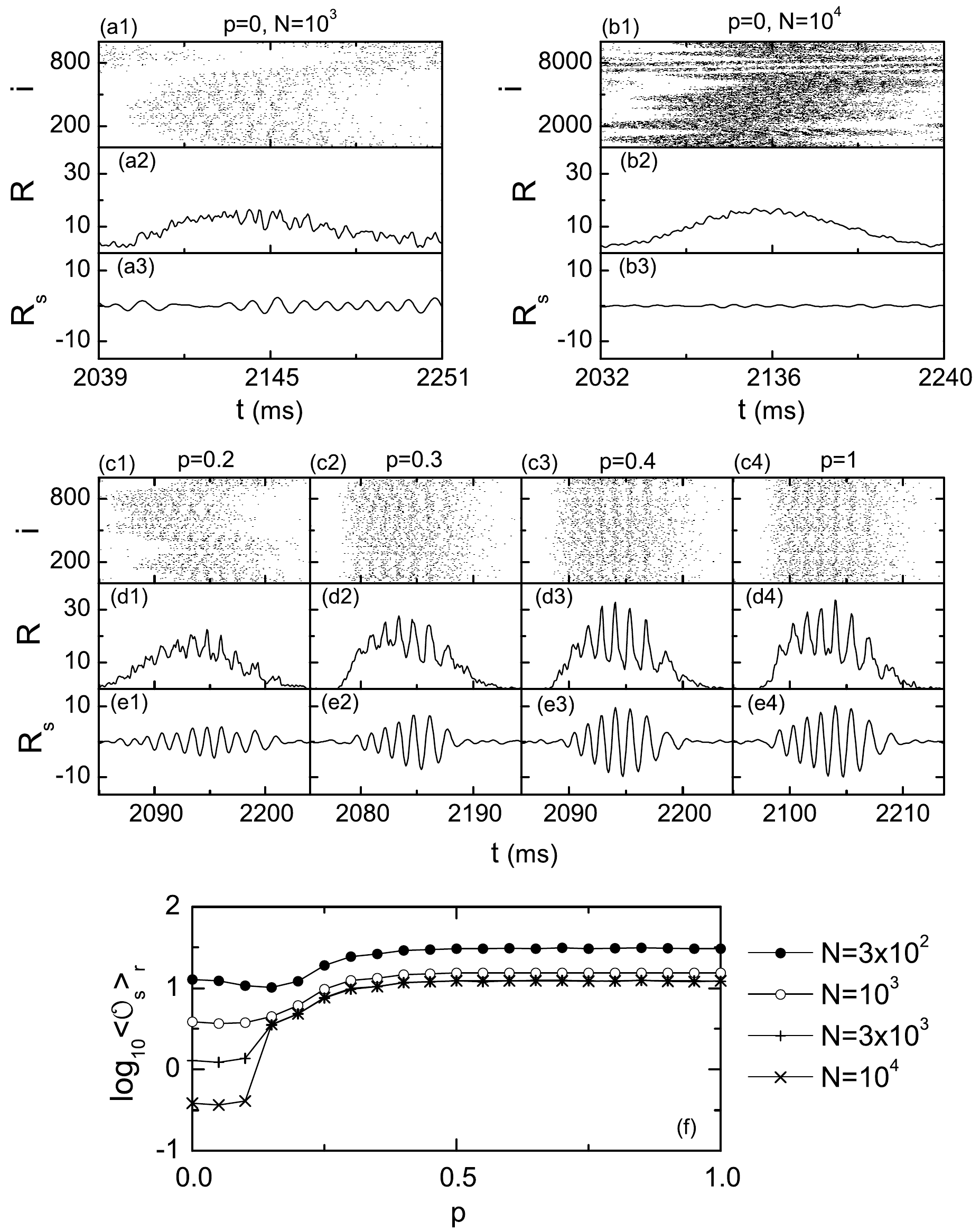}
\caption{Intraburst spiking transition with increasing $p$ for $J=0.6$ and $D=0.03$ in the small-world networks of $N$ $[=10^3$ except for the case of (f)] inhibitory subthreshold bursting HR neurons for $I_{DC}=1.25$ and $M_{syn}=100$. Intraburst spike unsynchronized state for $p=0$: plots of raster plot of spikes, IPFR kernel estimates $R(t)$, and band-pass filtered IPSR $R_s(t)$ [lower and higher cut-off frequencies of 30 Hz (high-pass filter) and 90 Hz (low-pass filter)] in the 1st global bursting cycle of the IPBR $R_b(t)$ (after the transient time of $2 \times 10^3$ ms) in (a1)-(a3) for $N = 10^3$ and in (b1)-(b3) for $N = 10^4$. The band width $h$ of the Gaussian kernel function is 1 ms for the IPFR kernel estimate $R(t)$. (c1)-(c4) Raster plots of neural spikes, (d1)-(d4) IPFR kernel estimates $R(t)$, and (e1)-(e4) band-pass filtered IPSR $R_s(t)$ [lower and higher cut-off frequencies of 30 Hz (high-pass filter) and 90 Hz (low-pass filter)] in the 1st global bursting cycle of the IPBR $R_b(t)$ (after the transient time of $2 \times 10^3$ ms) for varioue spike-synchronized cases of $p=0.2,$ 0.3, 0.4, and 1. (f) Plots of spiking order parameters ${\langle {\cal{O}}_s \rangle}_r$ [based on $R_s(t)$] versus $p$. For each $p$, we follow 100 bursting cycles in each realization, and obtain ${\langle {\cal{O}}_s \rangle}_r$ via average over 10 realizations.
}
\label{fig:ST}
\end{figure}

As in the above case of the route ``A'' we also study the population behaviors along the route ``B'' for $J=0.6$ in Fig.~\ref{fig:SD1}. The raster plots of spikes and the IPFR kernel estimates $R(t)$ are shown in Figs.~\ref{fig:PIB}(d1)-\ref{fig:PIB}(d6) and Figs.~\ref{fig:PIB}(e1)-\ref{fig:PIB}(e6), respectively. When passing a bursting threshold $D$ $(\simeq 0.017)$, a transition from unsynchronization [e.g., see Figs.~\ref{fig:PIB}(d1) and \ref{fig:PIB}(e1) for $D=0.015$] to burst synchronization [e.g., see Figs.~\ref{fig:PIB}(d2) and \ref{fig:PIB}(e2) for $D=0.019$] occurs. For the case of burst synchronization, bursting bands (composed of spikes and indicating population synchronization) appear successively in the raster plot, and the IPFR kernel estimate $R(t)$ shows a slow-wave oscillation with the slow bursting timescale $\tau_b \simeq 207$ ms. As $D$ is increased and passes another lower spiking threshold $D$ $(\simeq 0.021)$, in addition to burst synchronization [synchrony on the slow bursting timescale $\tau_b$ $(\simeq 207$ ms)], spike synchronization [synchrony on the fast spike timescale $\tau_s$ $(\simeq 16$ ms)] occurs, as shown in Figs.~\ref{fig:PIB}(d3) and \ref{fig:PIB}(e3) for $D=0.03$. For this complete synchronization (including both the burst and spike synchronizations) each bursting band consists of ``spiking stripes'' and the corresponding IPFR kernel estimate $R(t)$ exhibits a bursting activity [i.e., fast spikes appear on the slow wave in $R(t)$], as clearly shown in the magnified 1st bursting band of Fig.~\ref{fig:ST}(c4) and in the magnified 1st bursting cycle of $R(t)$ in Fig.~\ref{fig:ST}(d4). Unlike the case of the route ``A,'' fast intraburst spike synchronization occurs for $J> J^*_2$ $(\simeq 0.394)$, in addition to the slow burst synchronization. However, such fast intraburst spike synchronization disappears due to overlap of spiking stripes in the bursting bands when passing a higher spiking threshold $D$ $(\simeq 0.043$). Then, only the burst-synchronized states (without fast spike synchronization) appear, as shown in Figs.~\ref{fig:PIB}(d4) and \ref{fig:PIB}(e4) for $D=0.06$. Like the above case of the route ``A,'' with further increase in $D$ the bursting bands become smeared, and hence the degree of burst synchronization decreases [e.g., see Figs.~\ref{fig:PIB}(d5) and \ref{fig:PIB}(e5) for $D=0.1$]. Eventually, when passing another higher bursting threshold $D$ $(\simeq 0.127$), a transition to unsynchronization occurs due to overlap of bursting bands, as shown in Figs.~\ref{fig:PIB}(d6) and \ref{fig:PIB}(e6) for $D=0.14$. Furthermore, the bursting patterns of individual HR neurons are the same as those for the above case of the route ``A,'' as shown in the IBI histograms of Figs.~\ref{fig:PIB}(f1)-\ref{fig:PIB}(f6). For the case of burst synchronization multi-peaked IBI histograms appear, while such peaks disappear due to their merging in the IBI histograms for the case of unsynchronization.

Throughout this paper, we consider only the case where the bursting type of individual HR neurons is the fold-homoclinic square-wave bursting which is
just the bursting type of the single HR neuron \citep{Rinzel1,Rinzel2,Burst3}. Unlike the single case, the bursting types of individual HR neurons
depend on the coupling strength $J$, as shown in Figs.~\ref{fig:PIB}(g1)-\ref{fig:PIB}(g3) along the route ``C'' for $D=0.03$ in Fig.~\ref{fig:SD1}.
For $J=0.7$, the bursting type of individual HR neurons is still the square-wave bursting, while the bursting type for $J=1.1$ is the fold-Hopf tapering bursting \citep{Burst3}. For an intermediate value (e.g., $J=0.9$), a mixed type of square wave and tapering burstings appear (i.e., square-wave and tapering burstings alternate).

So far, we have studied noise-induced burst and spike synchronizations in the conventional Erd\"{o}s-Renyi random graph of inhibitory subthreshold bursting HR neurons. For random connectivity, the average path length is short due to appearance of long-range connections, and hence global efficiency of information transfer becomes high \citep{Eff1,Eff2}. On the other hand, unlike the regular lattice, the random network has poor clustering \citep{Sporns,Buz2}. However, real synaptic connectivity is known to have complex topology which is neither regular nor completely random \citep{CN6,Buz2,CN1,CN2,CN3,CN7,CN4,CN5,Sporns}. To study the effect of network structure on noise-induced burst and spike synchronizations, we consider the Watts-Strogatz model for small-world networks which interpolates between regular lattice and random graph via rewiring \citep{SWN1}. By varying the rewiring probability $p$ from local to long-range connection, we investigate the effect of small-world connectivity on emergence of noise-induced burst and spike synchronizations. We start with a directed regular ring lattice with $N$ subthreshold bursting HR neurons where each HR neuron is coupled to its first $M_{syn}$ neighbors ($M_{syn}/2$ on either side) via outward synapses, and rewire each outward connection at random with probability $p$ such that self-connections and duplicate connections are excluded. As in the above random case, we consider a sparse but connected network with a fixed value of $M_{syn}= 100$. Then, we can tune the network between regularity $(p=0)$ and randomness $(p=1)$; the case of $p=1$ corresponds to the above Erd\"{o}s-Renyi random graph. In this way, we investigate emergence of noise-induced population synchronization in the directed Watts-Strogatz small-world network of $N$ inhibitory subthreshold bursting HR neurons by varying the rewiring probability $p$.

The topological properties of the small-world connectivity has been well characterized in terms of the clustering coefficient (local property) and the average path length (global property) \citep{SWN1}. The clustering coefficient, denoting the cliquishness of a typical neighborhood in the network, characterizes the local efficiency of information transfer, while the average path length, representing the typical separation between two vertices in the network, characterizes the global efficiency of information transfer. The regular lattice for $p=0$ is highly clustered but large world where the average path length grows linearly with $N$, while the random graph for $p=1$ is poorly clustered but small world where the average path length grows logarithmically with $N$ \citep{SWN1}. As soon as $p$ increases from 0, the average path length decreases dramatically, which leads to occurrence of a small-world phenomenon which is popularized by the phrase of the ``six degrees of separation'' \citep{SDS1,SDS2}. However, during this dramatic drop in the average path length, the clustering coefficient remains almost constant at its value for the regular lattice. Consequently, for small $p$ small-world networks with short path length and high clustering emerge \citep{SWN1}.

\begin{figure}
\includegraphics[width=0.7\columnwidth]{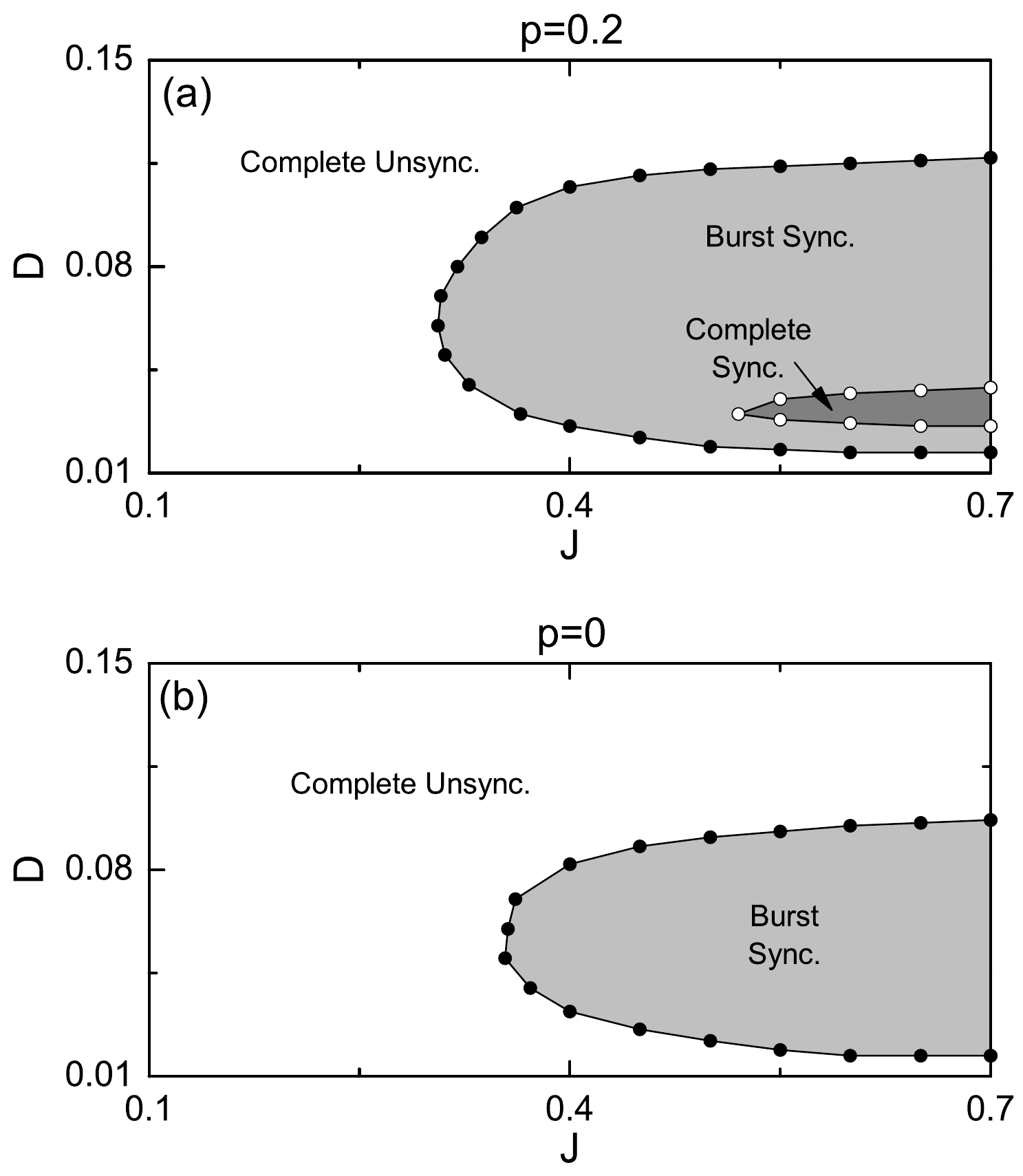}
\caption{State diagrams in the $J-D$ plane in (a) the Watts-Strogatz small-world network for $p=0.2$ and (b) the regular lattice
for $p=0$; each network consists of $N$ $(=10^3)$ inhibitory subthreshold bursting HR neurons for $I_{DC}=1.25$ and $M_{syn}=100$.
Complete synchronization (including both the burst and spike synchronizations) occur in the dark gray region, while
in the gray region only the burst synchronization appears.
}
\label{fig:SD2}
\end{figure}

We now investigate occurrence of noise-induced burst and spike synchronizations in the Watts-Strogatz small-world network of $N$ inhibitory subthreshold bursting HR neurons by decreasing the rewiring probability $p$ from 1 (random network). Figures \ref{fig:SD2}(a) and \ref{fig:SD2}(b) show the state diagrams in the $J-D$ plane for $p=0.2$ and 0, respectively. When comparing with the case of $p=1$ (random network) in Fig.~\ref{fig:SD1}, the gray region of slow burst synchronization decreases a little, while the dark-gray region of fast spike synchronization shrinks much more. As a result, only the burst synchronization (without fast spike synchronization) occurs in the regular lattice $(p=0)$. Unlike the case of the slow burst synchronization, more long-range connections are necessary for the emergence of fast spike synchronization. Hence, fast spike synchronization may occur only when the rewiring probability $p$ passes a (non-zero) critical value $p^*_c$ [e.g., $p_c^* \simeq 0.14$ for $J=0.6$ and $D=0.03$, as shown in Fig.~\ref{fig:ST}(f)].

\begin{figure}
\includegraphics[width=\columnwidth]{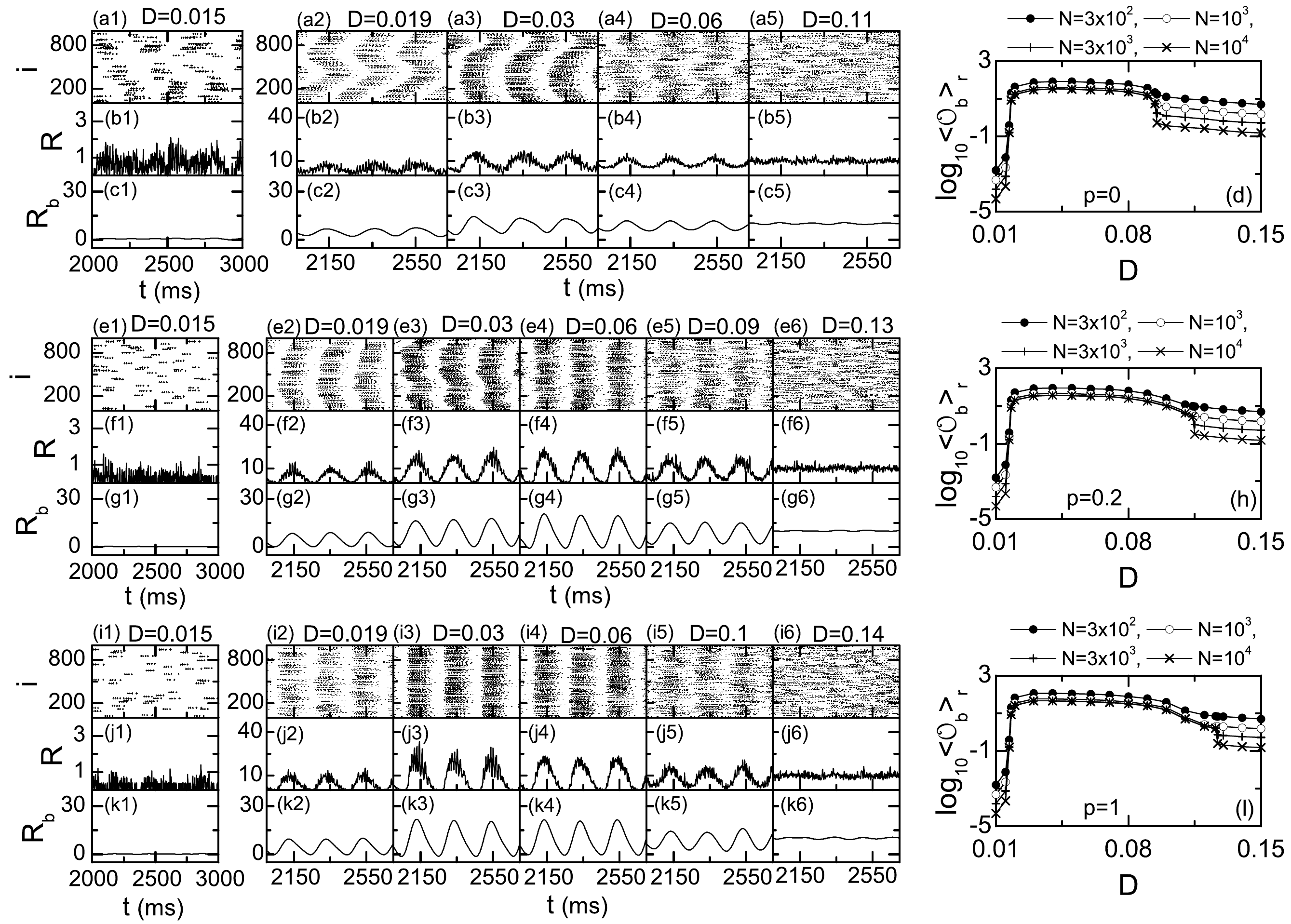}
\caption{Bursting transitions [based on the IPBR $R_b(t)$] with increasing $D$ for $J=0.6$ in the three cases of $p=0$ (regular lattice), $p=0.2$ (small-world network), and $p=1$ (random graph): each network consists of $N$ $[=10^3$ except for the cases of (d), (h), and (l)] inhibitory subthreshold bursting HR neurons for $I_{DC}=1.25$ and $M_{syn}=100$. Case of $p=0$: (a1)-(a5) raster plot of spikes, (b1)-(b5) IPFR kernel estimates $R(t)$, (c1)-(c5) low-pass filtered (cut-off frequency=10 Hz) IPBR $R_b(t)$, and (d) plots of bursting order parameters ${\langle {\cal{O}}_b \rangle}_r$ [based on $R_b(t)$] versus $D$. Case of $p=0.2$: (e1)-(e6) raster plot of spikes, (f1)-(f6) IPFR kernel estimates $R(t)$, (g1)-(g6) low-pass filtered (cut-off frequency=10 Hz) $R_b(t)$, and (h) plots of bursting order parameters ${\langle {\cal{O}}_b \rangle}_r$ versus $D$. Case of $p=1$: (i1)-(i6) raster plot of spikes, (j1)-(j6) IPFR kernel estimates $R(t)$, (k1)-(k6) low-pass filtered (cut-off frequency=10 Hz) $R_b(t)$, and (l) plots of bursting order parameters ${\langle {\cal{O}}_b \rangle}_r$ versus $D$. The band width $h$ of the Gaussian kernel function is 1 ms for the IPFR kernel estimate $R(t)$.
}
\label{fig:BT1}
\end{figure}

We first study bursting transitions (i.e., transitions to slow burst synchronization) with increasing $D$ for $J=0.6$ in the three cases of $p=0$ (regular lattice), 0.2 (small-world network), and 1 (random network). Figures \ref{fig:BT1}(a1)-\ref{fig:BT1}(a5) and \ref{fig:BT1}(b1)-\ref{fig:BT1}(b5) show the raster plots of spikes and the IPFR kernel estimate $R(t)$ for $p=0$. We note that the IPFR kernel estimate $R(t)$ is a population quantity describing the ``whole'' combined collective behaviors (including both the burst and spike synchronizations) of bursting neurons. For more clear investigation of burst synchronization, we separate the slow bursting timescale and the fast spiking timescale via frequency filtering, and decompose the IPFR kernel estimate $R(t)$ into the IPBR $R_b(t)$ and the IPSR $R_s(t)$. Through low-pass filtering of $R(t)$ with cut-off frequency of 10 Hz, we obtain the IPBR $R_b(t)$ (containing only the bursting behavior without spiking) for $p=0$ in Figs.~\ref{fig:BT1}(c1)-\ref{fig:BT1}(c5). Then, the mean square deviation of $R_b(t)$,
\begin{equation}
{\cal{O}}_b \equiv \overline{(R_b(t) - \overline{R_b(t)})^2},
 \label{eq:Border1}
\end{equation}
plays the role of a bursting order parameter ${\cal{O}}_b$, characterizing the bursting transition, where the overbar represents the time average \citep{Kim1}. The order parameter ${\cal{O}}_b$ may be regarded as a thermodynamic measure because it concerns just the macroscopic IPBR $R_b(t)$ without any consideration between $R_b(t)$ and microscopic individual burstings. Here, we discard the first time steps of a trajectory as transients for $2 \times 10^3$ ms, and then we compute ${\cal{O}}_b$ by following the trajectory for $10^4$ ms for each realization. We obtain $\langle{\cal{O}}_b\rangle_r$ via average over 10 realizations. In the thermodynamic limit of $N \rightarrow \infty$, the bursting order parameter $\langle{\cal{O}}_b\rangle_r$  approaches a non-zero (zero) limit value for the synchronized (unsynchronized) bursting state. Figure \ref{fig:BT1}(d) shows plots of the bursting order parameter $\langle{\cal{O}}_b\rangle_r$ versus $D$ for $p=0$. For $D^*_{b,l} (\simeq 0.017) < D < D^*_{b,h}$ $(\simeq 0.095$), synchronized bursting states appear because the values of $\langle{\cal{O}}_b\rangle_r$ become saturated to non-zero limit values in the thermodynamic limit of $N \rightarrow \infty$. However, for $D <D^*_{b,l}$ or $D> D^*_{b,h}$, the bursting order parameter $\langle{\cal{O}}_b\rangle_r$ tends to zero as $N \rightarrow \infty$, and hence unsynchronized bursting states exist. In the case of burst synchronization for $p=0$, the raster plot shows a zigzag pattern of inclined partial bursting bands of spikes [see Figs.~\ref{fig:BT1}(a2)- \ref{fig:BT1}(a4)], and the corresponding IPFR $R(t)$ and IPBR $R_b(t)$ exhibit slow-wave oscillations, as shown in Figs.~\ref{fig:BT1}(b2)-\ref{fig:BT1}(b4) and Figs.~\ref{fig:BT1}(c2)-\ref{fig:BT1}(c4). For $p=0$ the clustering coefficient is high, and hence inclined partial bursting bands (indicating local clustering of spikes) seem to appear. On the other hand, for the case of unsynchronization for $p=0$ the IPBR $R_b(t)$ becomes nearly stationary because spikes are scattered without forming zigzagged bursting bands in the raster plot, as shown in the cases of $D=0.015$ and 0.11. With increasing $p$, we also investigate another bursting transitions in terms of $\langle{\cal{O}}_b\rangle_r$. As shown in Figs.~\ref{fig:BT1}(d) ($p=0$), \ref{fig:BT1}(h) ($p=0.2$), and \ref{fig:BT1}(l) ($p=1$), the higher bursting threshold values $D^*_{b,h}$ increases with increase in $p$ (i.e., $D^*_{b,h}$ for $p=0$, 0.2, and 1 are 0.095, 0.115, and 0.127, respectively), while the lower bursting threshold $D^*_{b,l}$ $(\simeq 0.017)$ is nearly the same for the three cases of $p=0$, 0.2, and 1. In this way, as the rewiring probability $p$ is increased, the burst-synchronized range of $D$ increases gradually because the average synaptic path length (characterizing the global efficiency of information transfer) decreases due to appearance of long-range connections with increasing $p$. We also note that with increase in $p$ the zigzagness degree of bursting bands in the raster plots of spikes becomes reduced [e.g., compare Figs.~\ref{fig:BT1}(a2) ($p=0$), \ref{fig:BT1}(e2) ($p=0.2$), and \ref{fig:BT1}(i2) ($p=1$) for $D=0.019$] because the clustering coefficient (characterizing the local efficiency of information transfer) decreases as $p$ is increased.

\begin{figure}
\includegraphics[width=\columnwidth]{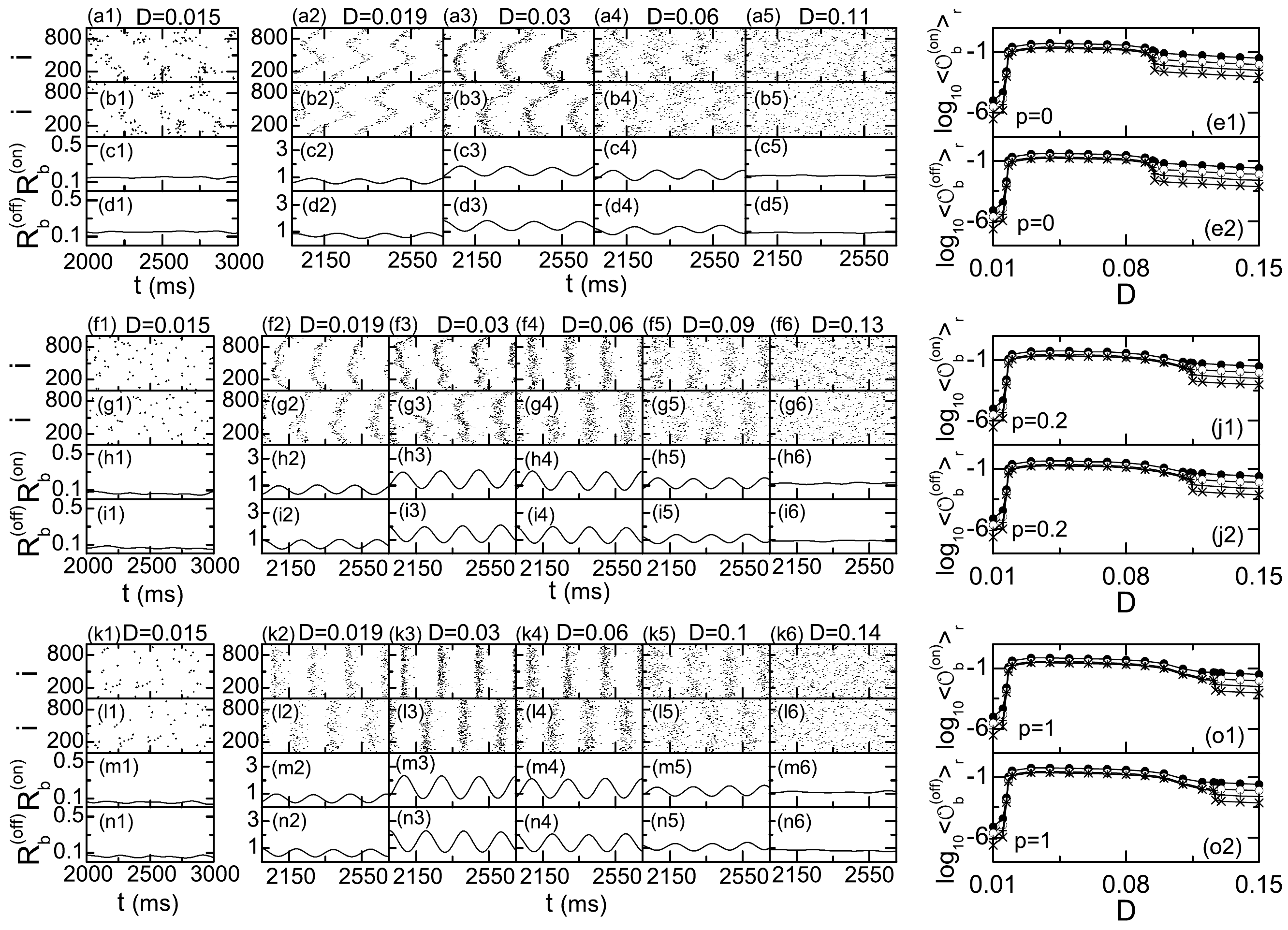}
\caption{Bursting transitions [based on the IPBRs $R_b^{(on)}(t)$ and $R_b^{(off)}(t)$] with increasing $D$ for $J=0.6$ in the three cases of $p=0$ (regular lattice), $p=0.2$ (small-world network), and $p=1$ (random graph): each network consists of $N$ $[=10^3$ except for the cases of the bursting order parameters, ${\langle {\cal{O}}_b^{(on)} \rangle}_r$ and ${\langle {\cal{O}}_b^{(off)} \rangle}_r$] inhibitory subthreshold bursting HR neurons for $I_{DC}=1.25$ and $M_{syn}=100$. Case of $p=0$: (a1)-(a5) raster plots of active phase (bursting) onset times, (b1)-(b5) raster plots of active phase (bursting) onset times, (c1)-(c5) IPBR kernel estimates $R_b^{(on)}(t)$, (d1)-(d5) IPBR kernel estimates $R_b^{(off)}(t)$, and plots of (e1) bursting order parameters ${\langle {\cal{O}}_b^{(on)} \rangle}_r$ [based on $R_b^{(on)}(t)$] and (e2) ${\langle {\cal{O}}_b^{(off)} \rangle}_r$ [based on $R_b^{(off)}(t)$] versus $D$. Case of $p=0.2$: (f1)-(f6) raster plot of active phase (bursting) onset times, (g1)-(g6) raster plot of active phase (bursting) onset times, (h1)-(h6) IPBR kernel estimates $R_b^{(on)}(t)$, (i1)-(i6) IPBR kernel estimates $R_b^{(off)}(t)$, and plots of (j1) bursting order parameters ${\langle {\cal{O}}_b^{(on)} \rangle}_r$ [based on $R_b^{(on)}(t)$] and (j2) ${\langle {\cal{O}}_b^{(off)} \rangle}_r$ [based on $R_b^{(off)}(t)$] versus $D$. Case of $p=1$: (k1)-(k6) raster plot of active phase (bursting) onset times, (l1)-(l6) raster plot of active phase (bursting) onset times, (m1)-(m6) IPBR kernel estimates $R_b^{(on)}(t)$, (n1)-(n6) IPBR kernel estimates $R_b^{(off)}(t)$, and plots of (o1) bursting order parameters ${\langle {\cal{O}}_b^{(on)} \rangle}_r$ [based on $R_b^{(on)}(t)$] and (o2) ${\langle {\cal{O}}_b^{(off)} \rangle}_r$ [based on $R_b^{(off)}(t)$] versus $D$. The symbols of the solid circles, open circles, pluses, and crosses used in the bursting order parameters, ${\langle {\cal{O}}_b^{(on)} \rangle}_r$ and ${\langle {\cal{O}}_b^{(off)} \rangle}_r$ represent $N= 3 \times 10^2$, $10^3$, $3 \times 10^3$, and $10^4$, respectively. The band width $h$ of the Gaussian kernel function is 50 ms for the IPBR kernel estimates $R_b^{(on)}(t)$ and $R_b^{(off)}(t)$.}
\label{fig:BT2}
\end{figure}

For more direct visualization of bursting behavior, we consider another raster plot of bursting onset or offset times [e.g., see the solid or open circles in Fig.~\ref{fig:Single}(b)], from which we can directly obtain the IPBR kernel estimate of band width $h=50$ ms, $R_b^{(on)}(t)$ or $R_b^{(off)}(t)$, without frequency filtering. Based on $R_b^{(on)}(t)$ and $R_b^{(off)}(t)$, we investigate bursting transitions with increasing $D$ for $J=0.6$ in the three cases of $p=0,$ 0.2, and 1. Figures \ref{fig:BT2}(a1)-\ref{fig:BT2}(a5) show the raster plots of the bursting onset times for $p=0$, while the raster plots of the bursting offset times are shown in Figs.~\ref{fig:BT2}(b1)-\ref{fig:BT2}(b5). From these raster plots of the bursting onset (offset) times, we obtain smooth IPBR kernel estimates, $R_b^{(on)}(t)$ [$R_b^{(off)}(t)$] in Figs.~\ref{fig:BT2}(c1)-\ref{fig:BT2}(c5) [\ref{fig:BT2}(d1)-\ref{fig:BT2}(d5)]. Then, the mean square deviations of $R_b^{(on)}(t)$ and $R_b^{(off)}(t)$,
\begin{equation}
{\cal{O}}_b^{(on)} \equiv \overline{(R_b^{(on)}(t) - \overline{R_b^{(on)}(t)})^2}~{\rm {and}}~
{\cal{O}}_b^{(off)} \equiv \overline{(R_b^{(off)}(t) - \overline{R_b^{(off)}(t)})^2},
 \label{eq:Border2}
\end{equation}
play another bursting order parameters which characterize the bursting transition \citep{Kim1}.
As in the the case of ${\cal{O}}_b$, we discard the first time steps of a trajectory as transients for $2 \times 10^3$ ms and then we compute  ${\cal{O}}_b^{(on)}$ and ${\cal{O}}_b^{(off)}$ by following the trajectory for $10^4$ ms for each realization. Thus, we obtain $\langle{\cal{O}}_b^{(on)}\rangle_r$ and $\langle{\cal{O}}_b^{(off)}\rangle_r$ via average over 10 realizations. Figures \ref{fig:BT2}(e1) and \ref{fig:BT2}(e2) show plots of the bursting order parameters $\langle{\cal{O}}_b^{(on)}\rangle_r$ and $\langle{\cal{O}}_b^{(off)}\rangle_r$ versus $D$ for $p=0$, respectively. Like the case of $\langle{\cal{O}}_b\rangle_r$, in the same region of $D^*_{b,l} (\simeq 0.017) < D < D^*_{b,h}$ $(\simeq 0.095$), synchronized bursting states exist because the values of $\langle{\cal{O}}_b^{(on)}\rangle_r$ and $\langle{\cal{O}}_b^{(off)}\rangle_r$ become saturated to non-zero limit values as $N \rightarrow \infty$. On the other hand, for $D <D^*_{b,l}$ or $D> D^*_{b,h}$, the bursting order parameters $\langle{\cal{O}}_b^{(on)}\rangle_r$ and $\langle{\cal{O}}_b^{(off)}\rangle_r$ tend to zero in the thermodynamic limit of $N \rightarrow \infty$, and hence unsynchronized bursting states appear. In this way, the bursting transition may also be well described in terms of the bursting order parameters  $\langle{\cal{O}}_b^{(on)}\rangle_r$ and $\langle{\cal{O}}_b^{(off)}\rangle_r$. In the case of burst synchronization for $p=0$, zigzagged bursting ``stripes,'' composed of bursting onset (offset) times, are formed in the raster plots of Figs.~\ref{fig:BT2}(a2)-\ref{fig:BT2}(a4) [Figs.~\ref{fig:BT2}(b2)-\ref{fig:BT2}(b4)]; the bursting onset and offset stripes are time-shifted [e.g., compare Figs.~\ref{fig:BT2}(a2) and \ref{fig:BT2}(b2) for $D=0.019$]. Since the clustering coefficient is high for $p=0$, zigzagged bursting onset and offset stripes (indicating local clustering of bursting onset and offset times) seem to appear. For this synchronous case, the corresponding IPBR kernel estimates, $R_b^{(on)}(t)$ and $R_b^{(off)}(t)$, show slow-wave oscillations with the same population bursting frequency $f_b$ $(\simeq 4.8$ Hz), as shown in Figs.~\ref{fig:BT2}(c2)-\ref{fig:BT2}(c4) and Figs.~\ref{fig:BT2}(d2)-\ref{fig:BT2}(d4), respectively, although they are phase-shifted [e.g., compare Figs.~\ref{fig:BT2}(c2) and \ref{fig:BT2}(d2) for $D=0.019$]. In terms of $\langle{\cal{O}}_b^{(on)}\rangle_r$ and $\langle{\cal{O}}_b^{(off)}\rangle_r$, we also investigate another bursting transitions with increasing $p$. Figures \ref{fig:BT2}(j1) and \ref{fig:BT2}(o1) [\ref{fig:BT2}(j2) and \ref{fig:BT2}(o2)] show plots of the bursting order parameter $\langle{\cal{O}}_b^{(on)}\rangle_r$ [$\langle{\cal{O}}_b^{(off)}\rangle_r$] versus $D$ for $p=0.2$ and 1, respectively. The burst-synchronized ranges of $D$ for $p=0.2$ and 1 are the same as those for the case of  $\langle{\cal{O}}_b\rangle_r$ [see Figs.~\ref{fig:BT1}(h) and \ref{fig:BT1}(l)], and they increase as $p$ is increased because the average synaptic path length (characterizing the global efficiency of information transfer) decreases due to appearance of long-range connections. Furthermore, with increase in $p$, the zigzagness degree of bursting onset and offset stripes in the raster plots becomes reduced [e.g., compare Figs.~\ref{fig:BT2}(a2) [\ref{fig:BT2}(b2)], \ref{fig:BT2}(f2) [\ref{fig:BT2}(g2)], and \ref{fig:BT2}(k2) [\ref{fig:BT2}(l2)] for $D=0.019$] because the clustering coefficient (characterizing the local efficiency of information transfer) decreases as $p$ is increased

In addition to the bursting transition, we also investigate spiking transitions (i.e., transitions to intraburst spike synchronization) of bursting HR neurons by varying the rewiring probability $p$ for $J=0.6$ and $D=0.03$. We first consider the case of $p=0$ (regular lattice) with long synaptic path length (corresponding to a large world). Figures \ref{fig:ST}(a1) and \ref{fig:ST}(a2) show the raster plot of intraburst spikes and the corresponding IPFR kernel estimate $R(t)$ during the 1st global bursting cycle of the IPBR $R_b(t)$ for $N=10^3$, respectively. As mentioned above, $R(t)$ exhibits the whole combined population behaviors including the burst and spike synchronizations with both the slow bursting and the fast spiking timescales. Hence, through band-pass filtering of $R(t)$ [with the lower and the higher cut-off frequencies of 30 Hz (high-pass filter) and 90 Hz (low-pass filer)], we obtain the IPSR $R_s(t)$, which is shown in Fig.~\ref{fig:ST}(a3). Then, the intraburst spike synchronization may be well described in terms of the IPSR $R_s(t)$. For the case of $N=10^3$, the IPFR $R(t)$ shows an explicit slow-wave oscillation, and hence population burst synchronization occurs for $p=0$. However, occurrence of intraburst spike synchronization cannot be clearly seen for $N=10^3$, because the IPSR $R_s(t)$ is composed of coherent parts with regular oscillations and incoherent parts with irregular fluctuations. For more clear investigation of spike synchronization, we also consider the case of $N=10^4$. Figures \ref{fig:ST}(b1)-\ref{fig:ST}(b3) show the raster plot of intraburst spikes, the IPFR kernel estimate $R(t)$, and the IPSR $R_s(t)$ for $N=10^4$, respectively. No ordered structure cannot be seen in the raster plot and the IPSR $R_s(t)$ is nearly stationary. Hence, the population state for $p=0$ seems to have no intraburst spike synchronization. However, as $p$ is increased, long-range short-cuts begin to appear, and hence characteristic synaptic path length becomes shorter. Consequently, for sufficiently large $p$ we expect emergence of intraburst spike synchronization because global efficiency of information transfer becomes better. Figures \ref{fig:ST}(c1)-\ref{fig:ST}(c4), \ref{fig:ST}(d1)-\ref{fig:ST}(d4), and \ref{fig:ST}(e1)-\ref{fig:ST}(e4) show the raster plots of intraburst spikes, the IPFRs $R(t)$, and the IPSRs $R_s(t)$ during the 1st global bursting cycle of the IPBR $R_b(t)$ for various synchronized cases of $p=0.2$, 0.3, 0.4, and 1, respectively. Clear spiking stripes (composed of intraburst spikes and indicating population spike synchronization) appear in the bursting band of the 1st global bursting cycle of the IPBR $R_b(t)$, and the IPFR kernel estimate $R(t)$ exhibits a bursting activity [i.e., fast spikes appear on a slow wave in $R(t)$] due to the complete synchronization (including both the burst and spike synchronizations). However, the band-pass filtered IPSR $R_s(t)$ shows only the fast spiking oscillations (without a slow wave) with the population spiking frequency $f_s$ $(\simeq 63$ Hz). We also characterize this intraburst spiking transition in terms of a spiking order parameter, based on $R_s(t)$. The mean square deviation of $R_s(t)$ in the $i$th global bursting cycle,
\begin{equation}
{\cal{O}}_s^{(i)} \equiv \overline{(R_s(t) - \overline{R_s(t)})^2},
\end{equation}
plays the role of a spiking order parameter ${\cal{O}}_s^{(i)}$ in the $i$th global bursting cycle of the IPBR $R_b(t)$. By averaging ${\cal{O}}_s^{(i)}$ over a sufficiently large number $N_b$ of global bursting cycles, we obtain the thermodynamic spiking order parameter:
\begin{equation}
{\cal{O}}_s =  {\frac {1} {N_b}} \sum_{i=1}^{N_b} {\cal{O}}_s^{(i)}.
\end{equation}
For each realization we follow $100$ bursting cycles, and obtain the spiking order parameter ${\langle {\cal{O}}_s \rangle}_r$ via average over 10 realizations. Figure \ref{fig:ST}(f) shows plots of ${\langle {\cal{O}}_s \rangle}_r$ versus $p$. When passing the spiking threshold value $p^*_c$ $(\simeq 0.14$), a transition to intraburst spike synchronization occurs because the values of ${\langle {\cal{O}}_s \rangle}_r$ become saturated to non-zero limit values as $N \rightarrow \infty$. Consequently, for $p>p^*_c$ synchronized spiking states exist because sufficient number of long-range short cuts for emergence of intraburst spike synchronization appear. In this way, the intraburst spiking transition may be well described in terms of the spiking  order parameter ${\langle {\cal{O}}_s \rangle}_r$.

\begin{figure}
\includegraphics[width=0.8\columnwidth]{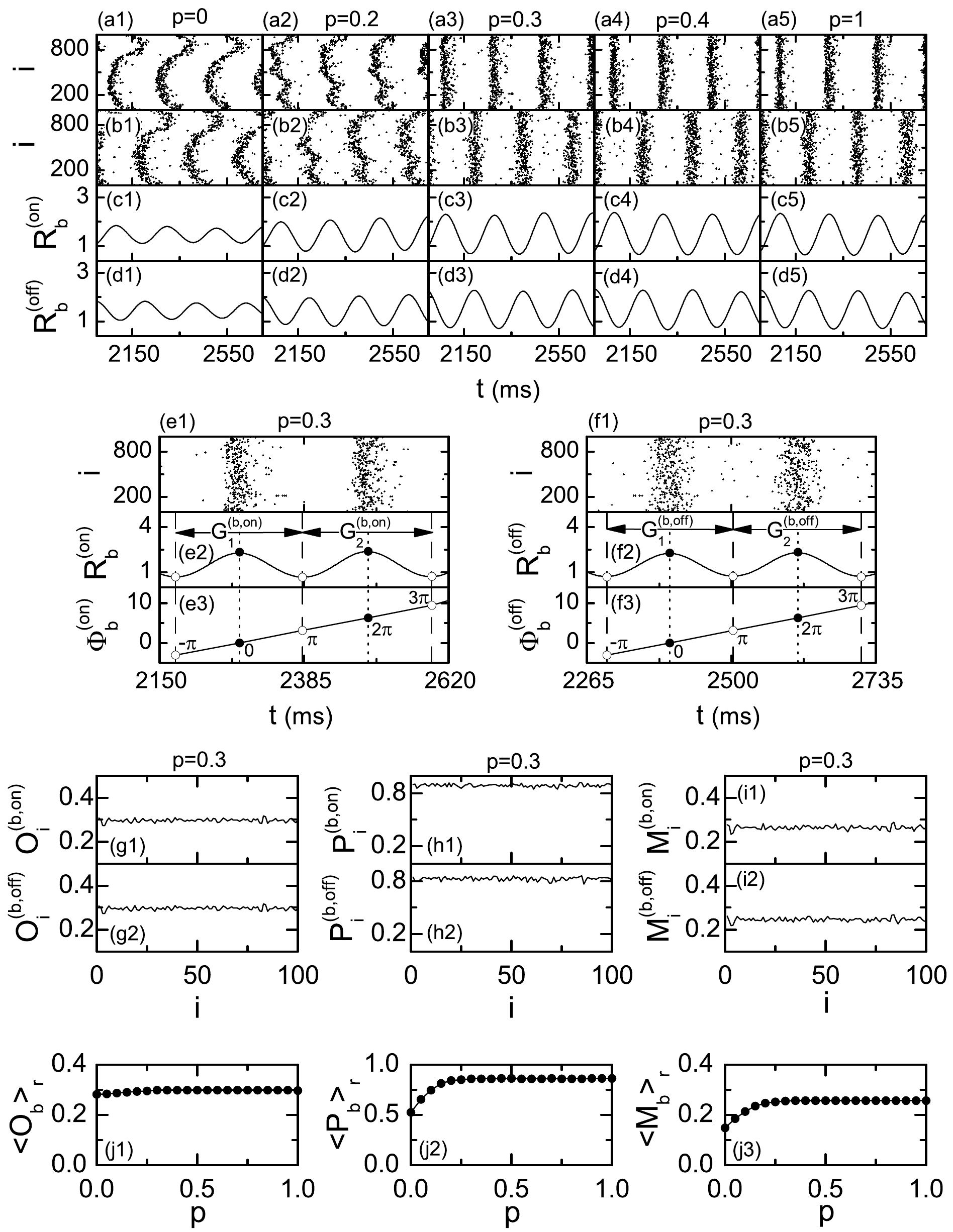}
\caption{Measurement of the degree of burst synchronization in terms of the statistical-mechanical bursting measure $M_b$ for $J=0.6$ and $D=0.03$ in the small-world networks of $N$ $(=10^3)$ inhibitory subthreshold bursting HR neurons for $I_{DC}=1.25$ and $M_{syn}=100$. (a1)-(a5) Raster plots of active phase (bursting) onset times, (b1)-(b5) raster plots of active phase (bursting) offset times, (c1)-(c5) IPBR kernel estimates $R_b^{(on)}(t)$, and (d1)-(d5) IPBR kernel estimates $R_b^{(off)}(t)$ for various values of $p$. For $p=0.3$, (e1) [(f1)] raster plot of active phase bursting onset (offset) times, (e2) [(f2)] IPBR kernel estimate $R_b^{(on)}(t)$ [$R_b^{(off)}(t)$], (e3) [(f3)] global bursting phase $\Phi_b^{(on)}(t)$ [$\Phi_b^{(off)}(t)$], and plots of (g1) [(g2)] $O_i^{(b,on)}$ [$O_i^{(b,off)}$] [occupation degree of bursting onset (offset) times in the $i$th global bursting onset (offset) cycle], (h1) [(h2)] $P_i^{(b,on)}$ [$P_i^{(b,off)}$] [pacing degree of bursting onset (offset) times in the $i$th global bursting onset (offset) cycle], and (i1) [(i2)] $M_i^{(b,on)}$ [$M_i^{(b,off)}$] [bursting measure in the $i$th global bursting onset (offset) cycle] versus $i$. In (e2)-(e3) and (f2)-(f3), vertical dashed and dotted lines represent the times at which local minima and maxima (denoted by open and solid circles) of $R_b^{(on)}(t)$ and $R_b^{(off)}(t)$ occur, respectively, and $G_i^{(b,on)}$ [$G_i^{(b,off)}$] $(i=1,2)$ denotes the ith global bursting onset (offset) cycle. Plots of (j1) ${\langle O_b \rangle}_r$ (average occupation degree of burstings), (j2) ${\langle P_b \rangle}_r$ (average pacing degree of burstings), and (j3) ${\langle M_b \rangle}_r$ (statistical-mechanical bursting measure) versus $p$. For each $p$, we follow 100 bursting cycles in each realization, and obtain ${\langle O_b \rangle}_r$, ${\langle P_b \rangle}_r$, and ${\langle M_b \rangle}_r$ via average over 10 realizations.
}
\label{fig:BM}
\end{figure}

From now on, we employ a statistical-mechanical bursting measure $M_b$, based on the IPBR kernel estimates $R_b^{(on)}(t)$ and $R_b^{(off)}(t)$ \citep{Kim1}, and measure the degree of burst synchronization by varying the rewiring probability $p$ for $J=0.6$ and $D=0.03$. As shown in Figs.~\ref{fig:BM}(a1)-\ref{fig:BM}(a5) [\ref{fig:BM}(b1)-\ref{fig:BM}(b5)], burst synchronization may be well visualized in the raster plots of bursting onset (offset) times. Clear bursting stripes (composed of bursting onset (offset) times and indicating population burst synchronization) appear in the raster plots. As $p$ is increased, the clustering coefficient (characterizing the local efficiency of information transfer) decreases, and hence the zigzagness degree of bursting onset and offset stripes becomes reduced. For this case of burst synchronization, both the IPBR kernel estimates $R_b^{(on)}(t)$ and $R_b^{(off)}(t)$ exhibit slow-wave oscillations, as shown in Figs.~\ref{fig:BM}(c1)-\ref{fig:BM}(c5) and Figs.~\ref{fig:BM}(d1)-\ref{fig:BM}(d5), respectively. As an example, we consider a synchronous bursting case of $p=0.3$. We measure the the degree of the burst synchronization seen in the raster plot of bursting onset (offset) times in Fig.~\ref{fig:BM}(e1)[\ref{fig:BM}(f1)] in terms of a statistical-mechanical bursting measure $M_b^{(on)}$ [$M_b^{(off)}$], based on $R_b^{(on)}(t)$ [$R_b^{(off)}(t)$], which is developed by considering the occupation pattern and the pacing pattern of the bursting onset (offset) times in the bursting stripes \citep{Kim1}. We first consider the raster plot of the bursting onset times. The bursting measure $M_i^{(b,on)}$ of the $i$th bursting onset stripe is defined by the product of the occupation degree $O_i^{(b,on)}$ of bursting onset times (representing the density of the $i$th bursting onset stripe) and the pacing degree $P_i^{(b,on)}$ of bursting onset times (denoting the smearing of the $i$th bursting onset stripe):
\begin{equation}
M_i^{(b,on)} = O_i^{(b,on)} \cdot P_i^{(b,on)}.
\label{eq:BM1}
\end{equation}
The occupation degree $O_i^{(b,on)}$ of bursting onset times in the $i$th bursting stripe is given by the fraction of HR neurons which exhibit burstings:
\begin{equation}
   O_i^{(b,on)} = \frac {N_i^{(b)}} {N},
\label{eq:Occu}
\end{equation}
where $N_i^{(b)}$ is the number of HR neurons which exhibit burstings in the $i$th bursting stripe. For the full occupation $O_i^{(b,on)}=1$, while for the partial occupation $O_i^{(b,on)}<1$. The pacing degree $P_i^{(b,on)}$ of bursting onset times in the $i$th bursting stripe can be determined in a statistical-mechanical way by taking into account their contributions to the macroscopic IPBR kernel estimate $R_b^{(on)}(t)$.
The IPBR kernel estimate $R_b^{(on)}(t)$ for $p=0.3$ is  shown in Fig.~\ref{fig:BM}(e2); local maxima and minima are represented by solid and open circles, respectively. Obviously, central maxima of $R_b^{(on)}(t)$ between neighboring left and right minima of $R_b^{(on)}(t)$ coincide with centers of bursting stripes in the raster plot.
The global bursting cycle starting from the left minimum of $R_b^{(on)}(t)$ which appears first after the transient time $(=2 \times 10^3$ ms) is regarded as the 1st one, which is denoted by $G_1^{(b,on)}$. The 2nd global bursting cycle $G_2^{(b,on)}$ begins from the next following right minimum of $G_1^{(b,on)}$, and so on. Then, we introduce an instantaneous global bursting phase $\Phi_b^{(on)}(t)$ of $R_b^{(on)}(t)$ via linear interpolation in the two successive subregions forming a global bursting cycle \citep{Kim1}, as shown in Fig.~\ref{fig:BM}(e3). The global bursting phase $\Phi_b^{(on)}(t)$ between the left minimum (corresponding to the beginning point of the $i$th global bursting cycle) and the central maximum is given by:
\begin{eqnarray}
\Phi_b^{(on)}(t) &=& 2\pi(i-3/2) + \pi \left(
\frac{t-t_i^{(on,min)}}{t_i^{(on,max)}-t_i^{(on,min)}} \right) \label{eq:Phi1} \\
& & {\rm~~ for~} ~t_i^{(on,min)} \leq  t < t_i^{(on,max)}
~~(i=1,2,3,\dots), \nonumber
\end{eqnarray}
and $\Phi_b^{(on)}(t)$ between the central maximum and the right minimum (corresponding to the beginning point of the $(i+1)$th global bursting cycle) is given by
\begin{eqnarray}
\Phi_b^{(on)}(t) &=& 2\pi(i-1) + \pi \left(
\frac{t-t_i^{(on,max)}}{t_{i+1}^{(on,min)}-t_i^{(on,max)}} \right) \label{eq:Phi2} \\
& & {\rm~~ for~} ~t_i^{(on,max)} \leq  t < t_{i+1}^{(on,min)}
~~(i=1,2,3,\dots), \nonumber
\end{eqnarray}
where $t_i^{(on,min)}$ is the beginning time of the $i$th global bursting cycle (i.e., the time at which the left minimum of $R_b^{(on)}(t)$ appears in the $i$th global bursting cycle) and $t_i^{(on,max)}$ is the time at which the maximum of $R_b^{(on)}(t)$ appears in the $i$th global bursting cycle. Then, the contribution of the $k$th microscopic bursting onset time in the $i$th bursting stripe occurring at the time $t_k^{(b,on)}$ to $R_b^{(on)}(t)$ is given by $\cos \Phi_k^{(b,on)}$, where $\Phi_k^{(b,on)}$ is the global bursting phase at the $k$th bursting onset time [i.e., $\Phi_k^{(b,on)} \equiv \Phi_b^{(on)}(t_k^{(b,on)})$]. A microscopic bursting onset time makes the most constructive (in-phase) contribution to $R_b^{(on)}(t)$ when the corresponding global phase $\Phi_k^{(b,on)}$ is $2 \pi n$ ($n=0,1,2, \dots$), while it makes the most destructive (anti-phase) contribution to $R_b^{(on)}(t)$ when $\Phi_k^{(b,on)}$ is $2 \pi (n-1/2)$. By averaging the contributions of all microscopic bursting onset times in the $i$th stripe to $R_b^{(on)}(t)$, we obtain the pacing degree of spikes in the $i$th stripe:
\begin{equation}
 P_i^{(b,on)} ={ \frac {1} {B_i^{(on)}}} \sum_{k=1}^{B_i^{(on)}} \cos \Phi_k^{(b,on)},
\label{eq:Pacing}
\end{equation}
where $B_i^{(on)}$ is the total number of microscopic bursting onset times in the $i$th bursting stripe. By averaging $M_i^{(b,on)}$ of Eq.~(\ref{eq:BM1}) over a sufficiently large number $N_b$ of bursting stripes, we obtain the statistical-mechanical bursting measure $M_b^{(on)}$, based on the IPSR kernel estimate $R_b^{(on)}(t)$:
\begin{equation}
M_b^{(on)} =  {\frac {1} {N_b}} \sum_{i=1}^{N_b} M_i^{(b,on)}.
\label{eq:BM2}
\end{equation}
For $p=0.3$ we follow $100$ bursting stripes and get $O_i^{(b,on)}$, $P_i^{(b,on)}$, and $M_i^{(b,on)}$ in each $i$th bursting stripe, which are shown in Figs.~\ref{fig:BM}(g1), \ref{fig:BM}(h1), and \ref{fig:BM}(i1), respectively. Due to sparse burstings of individual HR neurons, the average occupation degree $O_b^{(on)}$ (=${\langle O_i^{(b,on)} \rangle}_b \simeq 0.3)$, where ${\langle \cdots \rangle}_b$ denotes the average over bursting stripes, is small. Hence, only a fraction (about 3/10) of the total HR neurons fire burstings in each bursting stripe. On the other hand, the average pacing degree  $P_b^{(on)}$ (=${\langle P_i^{(b,on)} \rangle}_b \simeq 0.89)$ is large in contrast to $O_b^{(on)}$. Hence, the statistical-mechanical bursting measure $M_b^{(on)}$ (=${\langle M_i^{(b,on)} \rangle}_b$), representing the degree of burst synchronization seen in the raster plot of bursting onset times, is about 0.26. In this way, the statistical-mechanical bursting measure $M_b^{(on)}$ can be used effectively for measurement of the degree of burst synchronization because $M_b^{(on)}$ concerns the pacing degree as well as the occupation degree of bursting onset times in the bursting stripes of the raster plot.

In addition to the case of bursting onset times, we also measure the degree of burst synchronization between the bursting offset times. Figures \ref{fig:BM}(f1) and \ref{fig:BM}(f2) show the raster plot composed of two stripes of bursting offset times and the corresponding IPBR $R_b^{(off)}$ for $p=0.3$, respectively; the 1st and 2nd global bursting cycles, $G_1^{(b,off)}$ and $G_2^{(b,off)}$, are shown. Then, as in the case of $\Phi_b^{(on)}(t)$,  one can introduce an instantaneous global bursting phase $\Phi_b^{(off)}(t)$ of $R_b^{(off)}(t)$ via linear interpolation in the two successive subregions forming a global bursting cycle, which is shown in Fig.~\ref{fig:BM}(f3). Similarly to the case of bursting onset times, we also measure the degree of the burst synchronization seen in the raster plot of bursting offset times in terms of a statistical-mechanical bursting measure $M_b^{(off)}$, based on $R_b^{(off)}(t)$, by considering the occupation and the pacing patterns of the bursting offset times in the bursting stripes. The bursting measure $M_i^{(b,off)}$ in the $i$th bursting stripe also is defined by the product of the occupation degree $O_i^{(b,off)}$ of bursting offset times and the pacing degree $P_i^{(b,off)}$ of bursting offset times in the $i$th bursting stripe. We also follow $100$ bursting stripes and get $O_i^{(b,off)}$, $P_i^{(b,off)}$, and $M_i^{(b,off)}$ in each $i$th bursting stripe for $p=0.3$, which are shown in Figs.~\ref{fig:BM}(g2), \ref{fig:BM}(h2), and \ref{fig:BM}(i2), respectively. For this case of bursting offset times, $O_b^{(off)}$ (=${\langle O_i^{(b,off)} \rangle}_b) \simeq 0.3$, $P_b^{(off)}$ (=${\langle P_i^{(b,off)} \rangle}_b) \simeq 0.83$, and $M_b^{(off)}$ (=${\langle M_i^{(b,off)} \rangle}_b$) $\simeq 0.25$. The pacing degree of offset times $P_b^{(off)}$ is a little smaller than the pacing degree of the onset times ($P_b^{(on)} \simeq 0.89$), although the occupation degrees $(\simeq 0.3)$ of the onset and the offset times are the same. We take into consideration both cases of the onset and offset times equally and define the average occupation degree $O_b$, the average pacing degree $P_b$, and the statistical-mechanical bursting measure $M_b$ as follows:
\begin{equation}
 O_b = [O_b^{(on)} + O_b^{(off)}]/2,~~P_b = [P_b^{(on)} + P_b^{(off)}]/2,~~{\rm{and}}~~M_b = [M_b^{(on)} + M_b^{(off)}]/2.
\label{eq:BM3}
\end{equation}
By increasing the rewiring probability from $p=0$, we follow 100 bursting stripes in each realization and measure the degree of burst synchronization in terms of ${\langle O_b \rangle}_r$ (average occupation degree), ${\langle P_b \rangle}_r$ (average pacing degree), and ${\langle M_b \rangle}_r$ (statistical-mechanical bursting measure) via average over 10 realizations in the whole region of burst synchronization, and the results are shown in Figs.~\ref{fig:BM}(j1)-\ref{fig:BM}(j3). The average occupation degree ${\langle O_b \rangle}_r$ (denoting the average density of bursting stripes in the raster plot) is nearly the same (about 0.3), independently of $p$. On the other hand, with increasing $p$, the average pacing degree ${\langle P_b \rangle}_r$ (representing the average smearing of the bursting stripes in the raster plot) increases rapidly due to appearance of long-range connections. However, the value of ${\langle P_b \rangle}_r$ saturates for $p=p_{b,max}$ $(\sim 0.3)$ because long-range short-cuts which appear up to $p_{b,max}$ play sufficient role to get maximal degree of burst pacing. This saturation of the average pacing degree can be seen well in the raster plots of bursting onset times [see Figs.~\ref{fig:BM}(a1)-\ref{fig:BM}(a5)] and bursting offset times [see Figs.~\ref{fig:BM}(b1)-\ref{fig:BM}(b5)]. With increasing $p$ the zigzagness degree of bursting stripes in the raster plots becomes reduced, eventually for $p=p_{b,max}$ the raster plot becomes composed of vertical bursting stripes without zigzag, and then the pacing degree between bursting onset and offset times  becomes nearly the same. In the whole region of burst synchronization, $R_b^{(on)}$ and $R_b^{(off)}$ show slow-wave oscillations with the population bursting frequency $f_b \simeq 4.8$ Hz, independently of $p$. The amplitudes of the IPBR kernel estimates $R_b^{(on)}$ and $R_b^{(off)}$ also increase up to $p=p_{b,max}$, and then its value becomes saturated. The statistical-mechanical bursting measure ${\langle M_b \rangle}_r$ (taking into account both the occupation and the pacing degrees of bursting onset and offset times) also makes a rapid increase up to $p=p_{b,max}$, because ${\langle O_b \rangle}_r$ is nearly independent of $p$. ${\langle M_b \rangle}_r$ is nearly equal to $3 {\langle P_b \rangle}_r /10$ because of the sparse occupation (${\langle O_b \rangle}_r \simeq 3/10$). In this way, we characterize burst synchronization in terms of the statistical-mechanical bursting measure ${\langle M_b \rangle}_r$ in the whole region of burst synchronization, and find that ${\langle M_b \rangle}_r$  reflects the degree of burst synchronization seen in the raster plot of onset and offset times very well.

\begin{figure}
\includegraphics[width=\columnwidth]{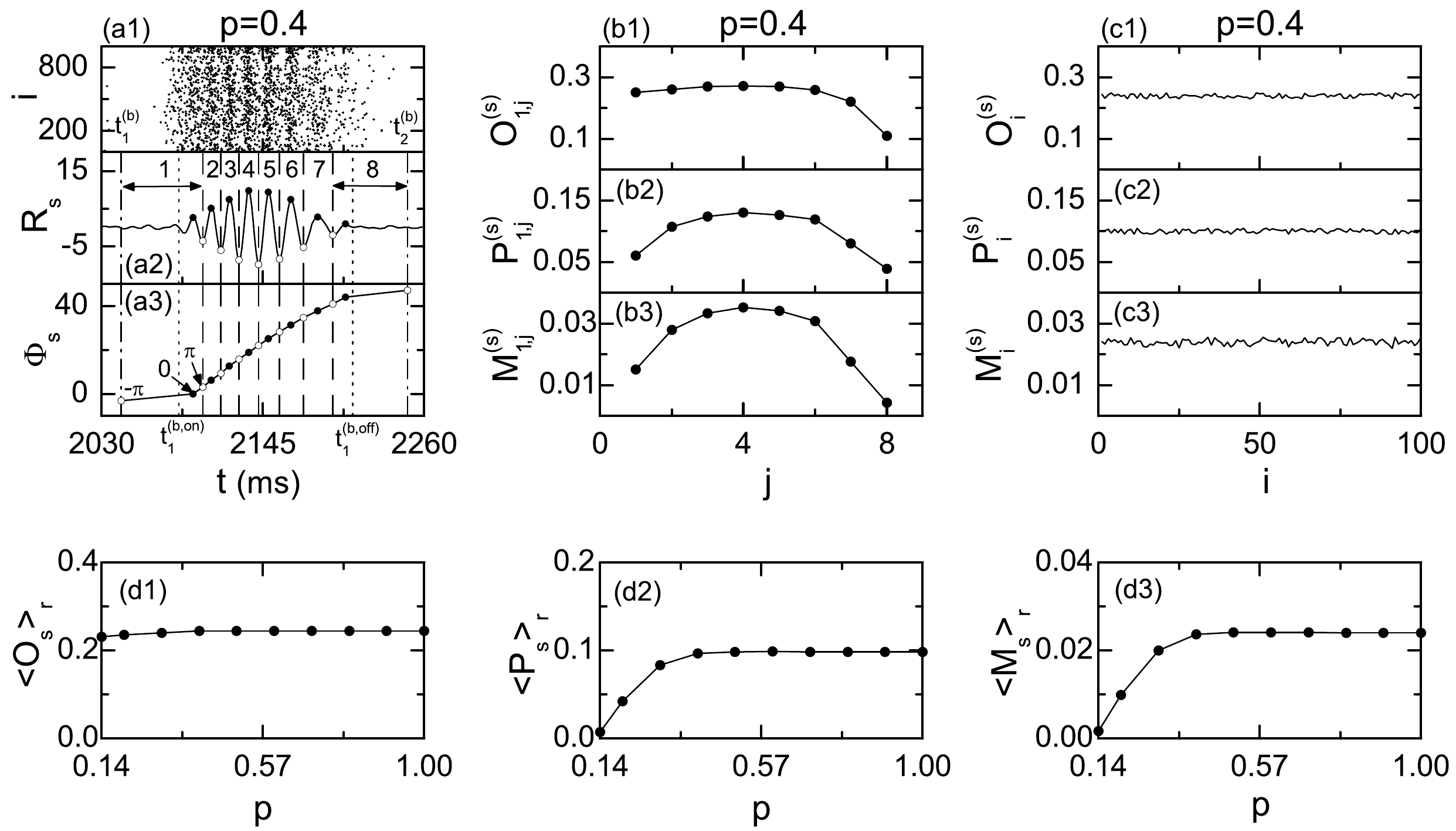}
\caption{Measurement of the degree of intraburst spike synchronization in terms of the statistical-mechanical spiking measure $M_s$ for $J=0.6$ and $D=0.03$ in the small-world networks of $N$ $(=10^3)$ inhibitory subthreshold bursting HR neurons for $I_{DC}=1.25$ and $M_{syn}=100$. (a1) Magnified raster plot of neural spikes, (a2) IPSR $R_s(t)$ (each integer $j$ $(=1, \dots, 8)$ represents the $j$th spiking cycle $G_{1,j}^{(s)}$), and (a3) global spiking phase $\Phi_s(t)$ in the 1st global bursting cycle of $R_b(t)$ [represented by the vertical  dash-dotted lines: $t_1^{(b)}$ (= 2044ms) $< t <$ $t_2^{(b)}$ (=2248 ms)] for $p=0.4$. The intraburst band in (a1) [denoted by the vertical dotted lines: $t_1^{(b,on)}$ (= 2085ms) $< t <$ $t_1^{(b,off)}$ (=2209 ms)], corresponding to the 1st global active phase, is composed of 8 smeared spiking stripes. Plots of (b1) $O_{1,j}^{(s)}$ (occupation degree of spikes), (b2) $P_{1,j}^{(s)}$ (pacing degree of spikes), and (b3) $M_{1,j}^{(s)}$ (spiking measure) in the $j$th spiking cycle $G_{1,j}^{(s)}$ of the 1st global bursting cycle of $R_b(t)$ versus $j$ for $p=0.4$. Plots of (c1) $O_i^{(s)}$ (occupation degree of spikes), (c2) $P_i^{(s)}$ (pacing degree of spikes), and (c3) $M_i^{(s)}$ (spiking measure) in the $i$th global bursting cycle versus $i$ for $p=0.4$. Measurement of the degree of intraburst spike synchronization: plots of (d1) ${\langle O_s \rangle}_r$ (average occupation degree of spikes), (d2) ${\langle P_s \rangle}_r$ (average pacing degree of spikes), and (d3) plot of ${\langle M_s \rangle}_r$ (average statistical-mechanical intraburst spiking measure) versus $p$. For each $p$, we follow 100 bursting cycles in each realization, and obtain ${\langle O_s \rangle}_r$, ${\langle P_s \rangle}_r$, and ${\langle M_s \rangle}_r$ via average over 10 realizations.
}
\label{fig:SM}
\end{figure}

Finally, We measure the degree of spike synchronization in terms of a statistical-mechanical spiking measure $M_s$, based on the IPSR $R_s(t)$. As shown in Figs.~\ref{fig:ST}(c1)-\ref{fig:ST}(c4), spike synchronization may be well visualized in the raster plot of spikes. For the synchronous spiking case, spiking stripes (composed of spikes and indicating intraburst spike synchronization) appear in the intraburst band of the raster plot. As an example, we consider a synchronous spiking case of $p=0.4$.
Figures \ref{fig:SM}(a1) and \ref{fig:SM}(a2) show a magnified raster plot of neural spikes and the IPSR $R_s(t)$, corresponding to the 1st global bursting cycle of the IPBR $R_b(t)$ [denoted by the vertical dash-dotted lines: $t_1^{(b)} (=2044~ {\rm ms}) < t < t_2^{(b)} (= 2248~ {\rm ms})$]. The intraburst band in Fig.~\ref{fig:SM}(a2) [represented by the vertical dotted lines: $t_1^{(b,on)} (=2085~ {\rm ms}) < t < t_2^{(b,off)} (= 2209~ {\rm ms})$], corresponding to the 1st global active phase, is composed of 8 smeared spiking stripes; $t_1^{(b,on)}$ (maximum of $R_b^{(on)}(t)$ in Fig.~\ref{fig:BM}(c4) within the 1st global bursting cycle) is the global active phase onset time, and $t_1^{(b,off)}$ (maximum of $R_b^{(off)}(t)$ in Fig.~\ref{fig:BM}(d4) within the 1st global bursting cycle) is the global active phase offset time. In the intraburst band (bounded by the dotted lines), the maxima (minima) of the IPSR $R_s(t)$ are denoted by solid (open) circles, and 8 global spiking cycles $G_{1,j}^{(s)}$ $(j=1, ... , 8)$ [denoted by the number $j$ in Fig.~\ref{fig:SM}(a2)] exist in the 1st global bursting cycle of $R_b(t)$. For $1<j<8$, each $j$th global spiking cycle $G_{1,j}^{(s)}$, containing the $j$th maximum of $R_s(t)$, begins at the left nearest-neighboring minimum of $R_s(t)$ and ends at the right nearest-neighboring minimum of $R_s(t)$, while for both extreme cases of $j=1$ and 8, $G_{1,1}^{(s)}$ begins at $t_1^{(b)}$ [the beginning time of the 1st global bursting cycle of $R_b(t)$] and $G_{1,8}^{(s)}$ ends at $t_2^{(b)}$ [the ending time of the 1st global bursting cycle of $R_b(t)$]. Then, as in the case of the global bursting phase $\Phi_b^{(on)}(t)$ [$\Phi_b^{(off)}(t)$] of $R_b^{(on)}(t)$ [$R_b^{(on)}(t)$], we introduce an instantaneous global spiking phase $\Phi_s(t)$ of $R_s(t)$ via linear interpolation in the two successive subregions (the left subregion joining the left beginning point and the central maximum and the right subregion joining the central maximum and the right ending point) forming a global spiking cycle [see Fig.~\ref{fig:SM}(a3)]. Similarly to the case of burst synchronization, we measure the degree of the intraburst spike synchronization seen in the raster plot in terms of a statistical-mechanical spiking measure, based on $R_s(t)$, by considering the occupation and the pacing patterns of spikes in the global spiking cycles. The spiking measure $M_{1,j}^{(s)}$ of the $j$th global spiking cycle in the 1st global bursting cycle is defined by the product of the occupation degree $O_{1,j}^{(s)}$ of spikes (denoting the density of spikes in the $j$th global spiking cycle) and the pacing degree $P_{1,j}^{(s)}$ of spikes (representing the smearing of spikes in the $j$th global spiking cycle). Plots of $O_{1,j}^{(s)}$, $P_{1,j}^{(s)}$, and $M_{1,j}^{(s)}$, are shown in Fig.~\ref{fig:SM}(b1)-\ref{fig:SM}(b3), respectively. For the 1st global bursting cycle, the spiking-averaged occupation degree $O_1^{(s)}$ (=${\langle O_{1,j}^{(s)} \rangle}_s$) $ \simeq 0.24$, the spiking-averaged pacing degree $P_1^{(s)}$ (=${\langle P_{1,j}^{(s)} \rangle}_s$) $ \simeq 0.1$, and the spiking-averaged statistical-mechanical spiking measure $M_1^{(s)}$ (=${\langle M_{1,j}^{(s)} \rangle}_s$) $ \simeq 0.024$, where ${\langle \cdots \rangle}_s$ represents the average over the spiking cycles. We also follow $100$ bursting cycles and get $O_i^{(s)}$, $P_i^{(s)}$, and $M_i^{(s)}$ in each $i$th global bursting cycle for $p=0.4$, which are shown in Figs.~\ref{fig:SM}(c1), \ref{fig:SM}(c2), and \ref{fig:SM}(c3), respectively. Then, through average over all bursting cycles, we obtain the bursting-averaged occupation degree $O_s$ (=${\langle O_i^{(s)} \rangle}_b \simeq 0.24)$, the bursting-averaged pacing degree $P_s$ (=${\langle P_i^{(s)} \rangle}_b \simeq 0.1)$, and the bursting-averaged statistical-mechanical spiking measure $M_s$ (=${\langle M_i^{(s)} \rangle}_b \simeq 0.024)$ for $p=0.4$. We note that $O_s$, $P_s$, and $M_s$ are obtained through double-averaging $[{\langle {\langle \cdots \rangle}_s \rangle}_b]$ over the spiking and bursting cycles. When compared with the bursting case of $O_b \simeq 0.3$ and $P_b \simeq 0.86$ for $p=0.4$, a fraction (about 4/5) of the HR neurons exhibiting the bursting active phases fire spikings in the spiking cycles, and the pacing degree of spikes ($P_s$) is about 12 percentage of the pacing degree of burstings ($P_b$). Consequently, the statistical-mechanical spiking measure ($M_s$) becomes only about 10 percentage of the statistical-mechanical bursting measure ($M_b$) for $p=0.4$ (i.e., the degree of the intraburst spike synchronization is much less than that of the burst synchronization). We increase the rewiring probability $p$ from 0 and repeat the above process to get $O_s$, $P_s$, and $M_s$ for multiple realizations. Thus, we obtain ${\langle O_s \rangle}_r$ (average occupation degree of spikes in the global spiking cycles), ${\langle P_s \rangle}_r$ (average pacing degree of spikes in the global spiking cycles), and ${\langle M_s \rangle}_r$ (average statistical-mechanical spiking measure in the global spiking cycles) through average over all realizations. For each realization, we follow 100 bursting cycles, and obtain ${\langle O_s \rangle}_r$, ${\langle P_s \rangle}_r$, and ${\langle M_s \rangle}_r$ via average over 10 realizations. Through these multiple-realization simulations, we measure the degree of intraburst spike synchronization in terms of ${\langle O_s \rangle}_r$, ${\langle P_s \rangle}_r$, and ${\langle M_s \rangle}_r$ in the whole region of spike synchronization [$p \geq   p^*_c (\simeq 0.14)$], which are shown in Figs.~\ref{fig:SM}(d1)-\ref{fig:SM}(d3), respectively. The average occupation degree ${\langle O_s \rangle}_r$ (denoting the average density of spiking stripes in the raster plot) is nearly the same (about 0.24), independently of $p$. On the other hand, with increasing $p$, the average pacing degree ${\langle P_s \rangle}_r$ (representing the average smearing of the spiking stripes in the raster plot) increases rapidly due to appearance of long-range connections. However, the value of ${\langle P_s \rangle}_r$ saturates for $p=p_{s,max}$ $(\sim 0.4)$ because long-range short-cuts which appear up to $p_{s,max}$ play sufficient role to get maximal degree of spike pacing. In this way, we characterize intraburst spike synchronization in terms of the average statistical-mechanical spiking measure ${\langle M_s \rangle}_r$ in the whole spike-synchronized region, and find that ${\langle M_s \rangle}_r$ reflects the degree of intraburst spike synchronization seen in the raster plot very well.

\section{Summary} \label{sec:SUM}
We have investigated the effect of network architecture on the noise-induced burst and spike synchronizations in an inhibitory population
of subthreshold bursting HR neurons. Noise-induced firing patterns of subthreshold bursting neurons, characterized by random skipping of bursts leading to a multi-modal IBI histogram, are in contrast to the deterministic firings for the suprathreshold case. For modeling the complex synaptic connectivity, we first employed the conventional Erd\"{o}s-Renyi random graph of subthreshold HR neurons, and studied occurrence of the noise-induced population synchronization by varying the synaptic inhibition strength $J$ and the noise intensity $D$. Thus, noise-induced burst and spike synchronizations have been found to occur in a synchronized region in the $J-D$ plane. However, real synaptic connections are known to be neither regular nor random. Hence, we considered the Watts-Strogatz model for small-world networks which interpolates between regular lattice and random network via rewiring. By varying the rewiring probability $p$, we have investigated the effect of small-world connectivity on emergence of noise-induced burst and spike synchronizations. With decreasing $p$ from 1 (random network) to 0 (regular lattice), the region of burst synchronization has been found to decrease slowly in the $J-D$ plane, while the region of spike synchronization has been found to shrink rapidly. Hence, complete synchronization (including both the burst and spike synchronizations) may occur only when $p$ is sufficiently large, whereas for small $p$ only burst synchronization (without spike synchronization) emerges because more long-range connections are necessary for the occurrence of fast spike synchronization. These burst and spike synchronizations may be well visualized in the raster plot of neural spikes which may be obtained in experiments. The IPFR kernel estimate $R(t)$, which is obtained from the raster plot of spikes, is a population quantity showing collective behaviors (including the burst and spike synchronizations) with both the slow bursting and the fast spiking timescales. Through frequency filtering, we have decomposed the IPFR kernel estimate $R(t)$ into the IPBR $R_b(t)$ and the IPSR $R_s(t)$, and characterized the noise-induced burst and spike synchronization transitions in terms of the bursting and spiking order parameters ${\cal {O}}_b$ and ${\cal {O}}_s$, based on $R_b(t)$ and $R_s(t)$, respectively. By varying $D$, we have investigated the noise-induced bursting transition in terms of ${\cal {O}}_b$ for a given $J$, and found that, with increasing the rewiring probability $p$ from 0 (regular lattice) the burst-synchronized range of $D$ increases gradually because long-range connections appear. For fixed $J$ and $D$, we have also studied the noise-induced spiking transition in terms of ${\cal {O}}_s$ by changing $p$. As $p$ passes a critical value $p^*_c$, a transition to spike synchronization has been found to occur in small-world networks, because sufficient number of long-range connections for occurrence of fast spike synchronization appear. We have also considered another raster plot of bursting onset or offset times for more direct visualization of bursting behavior. One can directly obtain the IPBR, $R_b^{(on)}(t)$ or $R_b^{(off)}(t)$, from this type of raster plot without frequency filtering. Then, we have characterized the bursting transitions in terms of another bursting order parameters, ${\cal {O}}_b^{(on)}$ and ${\cal {O}}_b^{(off)}$, based on $R_b^{(on)}(t)$ and $R_b^{(off)}(t)$. Furthermore, we have measured the degree of noise-induced burst synchronization seen in the raster plot of bursting onset or offset times in terms of a statistical-mechanical bursting measure $M_b$, introduced by considering the occupation and the pacing patterns of bursting onset or offset times in the raster plot. Similarly, we have also used a statistical-mechanical spiking measure $M_s$, based on $R_s$, and quantitatively measured the degree of the noise-induced intraburst spike synchronization. With increasing $p$, both the degrees of the noise-induced burst and spike synchronizations have been found to increase because more long-range connections appear. However, the degrees of the burst and spike synchronizations become saturated for their maximal values of $p$, $p_{b,max}$ $(\sim 0.3)$ and $p_{s,max}$
$(\sim 0.4)$, respectively because long-range short-cuts which appear up to the maximal values of $p$ play sufficient role to get maximal degrees of the burst and spike synchronizations.

\section*{Acknowledgments}
This research was supported by Basic Science Research Program through the National Research Foundation of Korea (NRF) funded by the Ministry of Education (Grant No. 2013057789).


\begin{thebibliography}{11}
\bibitem[Achard and Bullmore (2007)]{SW7} Achard S, Bullmore E (2007) Efficiency and cost of economical brain functional networks. PLoS Computational Biology 3:e17.
\bibitem[Bassett and Bullmore (2006)]{CN7} Bassett DS, Bullmore E (2006) Small-world brain networks. The Neuroscientist 12:512-523.
\bibitem[Batista et al. (2007)]{BSsync10} Batista CAS, Batista AM, de Pontes JAC, Viana RL, Lopes SR (2007) Chaotic phase synchronization in scale-free networks of bursting neurons. Physical Review E 76:016218.
\bibitem[Batista et al. (2012)]{BSsync11} Batista CAS, Lameu EL, Batista AM, Lopes SR, Pereira T, Zamora-Lopez G, Kurths J, Viana RL (2012) Phase synchronization of bursting neurons in clustered small-world networks. Physical Review E 86:016211.
\bibitem[B$\ddot{\rm o}$rgers and Kopell (2003)]{GABA1} B$\ddot{\rm o}$rgers C, Kopell N (2003) Synchronization in network of excitatory and inhibitory neurons with sparse, random connectivity. Neural Computation 15:509-538.
\bibitem[B$\ddot{\rm o}$rgers and Kopell (2005)]{GABA2} B$\ddot{\rm o}$rgers C, Kopell N (2005) Effects of noisy drive on rhythms in networks of excitatory and inhibitory neurons. Neural Computation 17:557-608.
\bibitem[Braun et al. (1994)]{Braun2} Braun HA, Wissing H, Sch\"{a}fer K, Hirsh MC (1994) Oscillation and noise determine signal transduction in shark multimodal sensory cells. Nature 367:270-273.
\bibitem[Brunel and Hakim (2008)]{Brunel} Brunel N, Hakim V (2008) Sparsely synchronized neuronal oscillations. Chaos 18:015113.
\bibitem[Bullmore and Sporns (2009)]{CN5} Bullmore E, Sporns O (2009) Complex brain networks: Graph-theoretical analysis of structural and functional systems. Nature Reviews Neuroscience 10:186-198.
\bibitem[Buzs$\acute{\rm a}$ki et al. (2004)]{Buz2} Buzs$\acute{\rm a}$ki G, Geisler C, Henze DA, Wang XJ (2004) Interneuron diversity series: circuit complexity and axon wiring economy of cortical interneurons. Trends in Neurosciences 27:186-193.
\bibitem[Chklovskii et al. (2004)]{CN1} Chklovskii DB, Mel BW, Svoboda K (2004) Cortical rewiring and information storage. Nature 431:782-788.
\bibitem[Coombes and Bressloff (2005)]{Burst1} Coombes S, Bressloff PC (eds) (2005) Bursting: the genesis of rhythm in the nervous system. World Scientific, Singapore. 
\bibitem[Dhamala et al. (2004)]{BSsync3} Dhamala M, Jirsa V, Ding M (2004) Transitions to synchrony in coupled bursting neurons. Physical Review Letters 92:028101.
\bibitem[Erd\"{o}s and Renyi (1959)]{ER} Erd\"{o}s P, Renyi A (1959) On random graph. Publicationes Mathematicae Debrecen 6:290-297.
\bibitem[Golomb and Rinzel (1994)]{KE1} Golomb D, Rinzel J (1994) Clustering in globally coupled inhibitory neurons. Physica D 72:259-282.
\bibitem[Guare (1990)]{SDS2} Guare J (1990) Six Degrees of Separation: A Play. Random House, New York.
\bibitem[Hindmarsh and Rose (1982)]{HR1} Hindmarsh JL, Rose RM (1982) A model of the nerve impulse using two first-order differential equations. Nature 296:162-164.
\bibitem[Hindmarsh and Rose (1984)]{HR2} Hindmarsh JL, Rose RM (1984) A model of neuronal bursting using three coupled first order differential equations. Proceedings of The Royal Society of London, Series B 221:87-102.
\bibitem[Hu and Zhou (2000)]{CR3} Hu B, Zhou C (2000) Phase synchronization in coupled nonidentical excitable systems and array-enhanced coherence resonance. Physical Review E 61:R1001-R1004.
\bibitem[Huber and Braun (2006)]{Braun1} Huber MT and Braun HA (2006) Stimulus-response curves of a neuronal model for noisy subthreshold oscillations and related spike generation. Physical Review E 73:041929.
\bibitem[Ivanchenko et al. (2004)]{BSsync4} Ivanchenko MV, Osipov GV, Shalfeev VD, Kurths J (2004) Phase synchronization in ensembles of bursting oscillators. Physical Review Letters 93:134101.
\bibitem[Izhikevich (2006).]{Burst2} Izhikevich EM (2006) Bursting. Scholarpedia 1(3):1300.
\bibitem[Izhikevich (2007)]{Burst3} Izhikevich EM (2007) Dynamical Systems in Neuroscience. MIT Press, Cambridge.
\bibitem[Kaiser and Hilgetag (2006)]{SW5} Kaiser M, Hilgetag CC (2006) Nonoptimal component placement, but short processing paths, due to long-distance projections in neural systems. PLoS Computational Biology 2:e95.
\bibitem[Kim and Lim (2014)]{Kim1} Kim SY, Lim W (2014) Thermodynamic and statistical-mechanical measures for characterization of the burst and spike synchronizations of bursting neurons. Submitted for publication in Journal of Neuroscience Methods. e-print: arXiv:1403.3994 [q-bio.NC].
\bibitem[Kwon and Moon (2002)]{SW3} Kwon O, Moon HT (2002) Coherence resonance in small-world networks of excitable cells. Physics Letters A 298:319-324.
\bibitem[Lago-Fern$\acute{\rm a}$ndez et al. (2000)]{SW2} Lago-Fern$\acute{\rm a}$ndez LF, Huerta R, Corbacho F, Sig$\ddot{\rm u}$enza JA (2000) Fast response and temporal coherent oscillations in small-world networks. Physical Review Letters 84:2758-2761.
\bibitem[Lameu et al. (2012)]{BSsync12} Lameu EL, Batista CAS, Batista AM, Larosz K, Viana RL, Lopes SR, Kurths J (2012) Suppression of bursting synchronization in clustered scale-free (rich-club) neural networks. Chaos 22:043149.
\bibitem[Larimer and Strowbridge (2008)]{CN4} Larimer P, Strowbridge BW (2008) Nonrandom local circuits in the dentate gyrus. Journal of Neuroscience 28:12212-12223.
\bibitem[Latora and Marchiori (2001)]{Eff1} Latora V, Marchiori M (2001) Efficient behavior of small-world networks. Physical Review Letters 87:198701.
\bibitem[Latora and Marchiori (2003)]{Eff2} Latora V, Marchiori M (2003). Economic small-world behavior in weighted networks. The European Physical Journal B 32:249-263.
\bibitem[Liang et al. (2009)]{HR4} Liang X, Tang M, Dhamala M, Liu Z (2009) Phase synchronization of inhibitory bursting neurons induced by distributed time delays in chemical coupling. Physical Review E 80:066202.
\bibitem[Lim and Kim (2011)]{Kim2} Lim W, Kim SY (2011) Statistical-mechanical measure of stochastic spiking coherence in a population of inhibitory subthreshold neuron. Journal of Computational Neuroscience 31:667-677.
\bibitem[Lizier et al. (2011)]{SW13} Lizier JT, Pritam S, Prokopenko M (2011) Information dynamics in small-world Boolean networks. Artificial Life 17:293-314.
\bibitem[Longtin (1997)]{Longtin2} Longtin A (1997). Autonomous stochastic resonance in bursting neurons. Physical Review E 55:868-876.
\bibitem[Longtin and Hinzer (1996)]{Longtin1} Longtin A, Hinzer K (1996) Encoding with bursting, subthreshold oscillations, and noise in mammalian cold receptors. Neural Computation 8:217-255.
\bibitem[Milgram (1967)]{SDS1} Milgram S (1967) The small-world problem. Psychology Today 1:61-67.
\bibitem[Neiman (2007)]{CR1} Neiman A (2007) Coherence resonance. Scholarpedia 2(11):1442.
\bibitem[Omelchenko et al. (2010)]{Burstsync2} Omelchenko I, Rosenblum M, Pikovsky A (2010) Synchronization of slow-fast systems. The European Physical Journal Special Topics 191:3-14.
\bibitem[Ozer et al. (2009)]{SW11} Ozer M, Perc M, Uzuntarla M (2009) Stochastic resonance on Newman-Watts networks of Hodgkin-Huxley neurons with local periodic driving. Physics Letters A 373:964-968.
\bibitem[Pereira et al. (2007)]{BSsync5} Pereira T, Baptista M, Kurths J (2007) Multi-time-scale synchronization and information processing in bursting neuron networks. The European Physical Journal Special Topics 146:155-168.
\bibitem[Riecke et al. (2007)]{SW6} Riecke H, Roxin A, Madruga S, Solla S (2007) Multiple attractors, long chaotic transients, and failure in small-world networks of excitable neurons. Chaos 17:026110.
\bibitem[Rinzel (1985)]{Rinzel1} Rinzel J (1985) Bursting oscillations in an excitable membrane model. In: Sleeman BD, Jarvis RJ (eds) Ordinary and Partial Differential Equations. Lecture Notes in Mathematics, Vol. 1151. Springer, Berlin, pp.~304-316.
\bibitem[Rinzel (1987)]{Rinzel2} Rinzel J (1987) A formal classiﬁcation of bursting mechanisms in excitable systems. In: Teramoto E, Yamaguti M (eds) Mathematical Topics in Population Biology, Morphogenesis, and Neurosciences. Lecture Notes in Biomathematics, Vol. 71. Springer-Verlag, Berlin, pp.~267-281.
\bibitem[Rose and Hindmarsh (1985)]{HR3} Rose RM, Hindmarsh JL (1985) A model of a thalamic neuron. Proceedings of The Royal Society of London, Series B 225:161-193.
\bibitem[Roxin et al. (2004)]{SW4} Roxin A, Riecke H, Solla SA (2004) Self-sustained activity in a small-world network of excitable neurons. Physical Review Letters 92:198101.
\bibitem[Rubin (2007)]{Burstsync1} Rubin JE (2007) Burst synchronization. Scholarpedia 2(10):1666.
\bibitem[San Miguel and Toral (2000)]{SDE} San Miguel M, Toral R (2000) Stochastic effects in physical systems. In: Martinez J, Tiemann R, Tirapegui E (eds) Instabilities and Nonequilibrium Structures VI. Kluwer Academic Publisher, Dordrecht, pp. 35-130.
\bibitem[Shanahan (2008)]{SW10} Shanahan M (2008) Dynamical complexity in small-world networks of spiking neurons. Physical Review E 78:041924.
\bibitem[Shi and Lu (2005)]{BSsync9} Shi X, Lu Q (2005) Firing patterns and complete synchronization of coupled Hindmarsh-Rose neurons. Chinese Physics 14:77-85.
\bibitem[Shi and Lu (2009)]{BSsync8} Shi X, Lu Q (2009) Burst synchronization of electrically and chemically coupled map-based neurons. Physica A 388:2410-2419.
\bibitem[Shimazaki and Shinomoto (2010)]{Kernel} Shimazaki H, Shinomoto S (2010) Kernel band width optimization in spike rate estimation. Journal of Computational Neuroscience 29:171-182.
\bibitem[Shinohara et al. (2002)]{CR6} Shinohara Y, Kanamaru T, Suzuki H, Horita T, Aihara K (2002) Array-enhanced coherence resonance and forced dynamics in coupled FitzHugh-Nagumo neurons with noise. Physical Review E 65:051906.
\bibitem[Song et al. (2005)]{CN2} Song S, Sj$\ddot{\rm o}$str$\ddot{\rm o}$m PJ, Reigl M, Nelson S, Chklovskii DB (2005) Highly nonrandom features of synaptic connectivity in local cortical circuits. PLoS Biology 3:e68.
\bibitem[Sporns (2011)]{Sporns} Sporns O (2011) Networks of the Brain. MIT Press, Cambridge.
\bibitem[Sporns and Honey (2006)]{CN3} Sporns O, Honey CJ (2006) Small worlds inside big brains, Proceedings of the National Academy of Sciences of the United States of America 103:19219-19220.
\bibitem[Sporns et al. (2000)]{CN6} Sporns O, Tononi G, Edelman GM (2000) Theoretical neuroanatomy: Relating anatomical and functional connectivity in graphs and cortical connection matrices. Cerebral Cortex 10:127-141.
\bibitem[Strogatz (2001)]{SWN2} Strogatz SH (2001) Exploring complex networks. Nature 410:268-276.
\bibitem[Sun et al. (2011)]{BSsync1} Sun X, Lei J, Perc M, Kurths J, Chen G (2011) Burst synchronization transitions in a neuronal network of subnetworks. Chaos 21:016110.
\bibitem[Tanaka et al. (2006)]{BSsync7} Tanaka G, Ibarz B, Sanjuan MA, Aihara K (2006) Synchronization and propagation of bursts in networks of coupled map neurons. Chaos 16:013113.
\bibitem[van Vreeswijk and Hansel (2001)]{BSsync2} van Vreeswijk C, Hansel D (2001) Patterns of synchrony in neural networks with adaptation. Neural Computation 13:959-992.
\bibitem[Wang (2010)]{Wang} Wang XJ (2010) Neurophysiological and computational principles of cortical rhythms in cognition. Physiological Reviews 90:1195-1268.
\bibitem[Wang and Buzs$\acute{\rm a}$ki (1996)]{KE2} Wang XJ, Buzs$\acute{\rm a}$ki G (1996) Gamma oscillations by synaptic inhibition in a hippocampal interneuronal network. Journal of Neuroscience 16:6402-6413.
\bibitem[Wang et al. (2000)]{CR2} Wang Y, Chik DTW, Wang ZD (2000) Coherence resonance and noise-induced synchronization in globally coupled Hodgkin-Huxley neurons. Physical Review E 61:740-746.
\bibitem[Wang et al. (2008)]{SW9} Wang Q, Duan Z, Perc M, Chen G (2008) Synchronization transitions on small-world neuronal networks: Effects of information transmission delay and rewiring probability. Europhysics Letters 83:50008.
\bibitem[Wang et al. (2010)]{SW12} Wang Q, Perc M, Duan Z, Chen G (2010) Impact of delays and rewiring on the dynamics of small-world neuronal networks with two types of coupling. Physica A 389:3299-3306.
\bibitem[Watts (2003)]{SWN3} Watts DJ (2003) Small Worlds: The Dynamics of Networks Between Order and Randomness. Princeton University Press.
\bibitem[Watts and Strogatz (1998)]{SWN1} Watts DJ, Strogatz SH (1998) Collective dynamics of 'small-world' networks. Nature 393:440-442.
\bibitem[Yu et al. (2008)]{SW8} Yu S, Huang D, Singer W, Nikolie D (2008) A small world of neuronal synchrony, Cerebral Cortex 18:2891-2901.
\bibitem[Yu et al. (2011)]{BSsync6} Yu H, Wang J, Deng B, Wei X, Wong YK, Chan WL, Tsang KM, Yu Z (2011) Chaotic phase synchronization in small world networks of bursting neurons. Chaos 21:013127.
\bibitem[Zhou and Kurths (2002)]{CR5} Zhou C, Kurths J (2002) Spatiotemporal coherence resonance of phase synchronization in weakly coupled chaotic oscillators. Physical Review E 65:040101.
\bibitem[Zhou et al. (2001)]{CR4} Zhou C, Kurths J, Hu B (2001) Array-enhanced coherence resonance: nontrivial effects of heterogeneity and spatial independence of noise. Physical Review Letters 87:098101.

\end{thebibliography}
\end{document}